\documentclass[onecolumn,floatfix,superscriptaddress,a4paper,nofootinbib,preprint]{revtex4}
\usepackage{epsfig}%
\usepackage{latexsym}
\usepackage{xspace}
\usepackage{hyperref}
\usepackage[utf8]{inputenc}
\usepackage{indentfirst}
\usepackage{enumerate}

\usepackage{gensymb}
\usepackage{amsmath}
\usepackage{amssymb}
\usepackage[english]{babel}
\usepackage{url}

\newcommand{\eq}[1]{\begin{align} #1 \end{align}}

\begin{document}

\title{Backward nucleon production by heavy baryonic resonances \\ in proton-nucleus collisions}

\author{Oleksandra~Panova}
\affiliation{Taras Shevchenko National University of Kyiv, Kyiv, Ukraine}

\author{Anton~Motornenko}
\affiliation{
Institut f\"ur Theoretische Physik,
Goethe Universit\"at, D-60438 Frankfurt am Main, Germany}
\affiliation{Frankfurt Institute for Advanced Studies, Giersch Science Center,
D-60438, Frankfurt am Main, Germany}

\author{Mark~I.~Gorenstein}
\affiliation{Bogolyubov Institute for Theoretical Physics, Kyiv, Ukraine}
\affiliation{Frankfurt Institute for Advanced Studies, Giersch Science Center,
D-60438, Frankfurt am Main, Germany}

\author{Jan~Steinheimer}
\affiliation{Frankfurt Institute for Advanced Studies, Giersch Science Center,
D-60438, Frankfurt am Main, Germany}

\author{Horst~Stoecker}
\affiliation{
Institut f\"ur Theoretische Physik,
Goethe Universit\"at, D-60438 Frankfurt am Main, Germany}
\affiliation{Frankfurt Institute for Advanced Studies, Giersch Science Center,
D-60438, Frankfurt am Main, Germany}
\affiliation{GSI Helmholtzzentrum f\"ur Schwerionenforschung GmbH, D-64291 Darmstadt,
Germany}

\begin{abstract}
The production of backward nucleons,  $N(180\degree)$, at $180\degree$ in the nuclear target rest frame in proton-nucleus (p~+~$A$) collisions is studied. 
The backward nucleons appearing outside of the kinematically allowed range of proton-nucleon (p~+~$N$)
reactions are shown to be due to secondary reactions of heavy baryonic resonances produced inside the nucleus. Baryonic resonances $R$ created in primary p~+~$N$ reactions can change their masses and momenta due to successive collisions $R+N\rightarrow R +N $  with other nuclear nucleons. Two distinct mechanisms and kinematic restrictions are studied: the reaction $R+N\rightarrow N(180\degree)+N$ and the resonance decay $R\rightarrow N(180\degree)+\pi$. Simulations of p~+~$A$ collisions using the Ultra-relativistic Quantum Molecular Dynamics 
model support these mechanisms and are consistent with  available data on proton backward production.

\vspace{0.5cm}
    
\end{abstract}

\maketitle


\section{Introduction}\label{sec -intr}

A large fraction of the particles produced in inelastic nucleon-nucleon ($N+N$) collisions 
appears from the decays of meson and baryon resonances,  e.g., from $\Delta$ and $N^*$ isobars, which decay subsequently into a stable baryon and one or more mesons, dileptons or photons. The large number of resonance states
and their large decay widths have led Rolf Hagedorn to postulate that the resonance mass spectrum behaves as a continuous exponentially increasing function~\cite{Hagedorn:1965st}. This is experimentally indeed found up to masses of approximately 3 GeV~\cite{Tanabashi:2018oca}, larger masses are not easily identified.
An exponentially increasing mass spectrum $\rho(m)\sim\exp(m/T_{\rm H})$ of the hadronic states at $m\rightarrow \infty$ leads to the limiting temperature $T=T_{\rm H}$ for strongly interacting matter.
Later theoretical suggestions~\cite{Hagedorn:1980kb,Gorenstein:1981fa,Stoecker:1981za} transformed the concept of a limiting temperature to the concept of a temperature of a phase transition or a crossover to a new high-temperature state -- the quark-gluon plasma. The crossover model with Hagedorn states does indeed describe the lattice QCD data~\cite{Vovchenko:2018eod}. Microscopic transport models of high energy collisions~\cite{Bass:1998ca,Bleicher:1999xi,Cassing:2008sv,Cassing:2009vt,Belkacem:1998gy} model the Hagedorn states by string excitations~\cite{Andersson:1983ia} or include them directly~\cite{Beitel:2014kza}.

Experimentally produced nucleons emitted in proton-nucleus (p~+~$A$) collisions have been observed in backward direction, at $180\degree$, in the nuclear target rest frame (these nucleons will be further denoted as $N(180\degree)$ and named as {\it backward} nucleons). Note, that the backward nucleons appear in a kinematic region forbidden in binary proton-proton (p~+~p) reactions. The production of backward nucleons and mesons was observed experimentally \cite{Baldin:1973pt,Baldin:1974sh} and is referred to as a 'cumulative effect' since
two or more nuclear nucleons should be involved in this process.   
Several models  were proposed to explain the data.
One group of models suggests that an extension of the kinematic limit of nucleon-nucleon ($N+N$) collision is possible in p~+~$A$ reactions due to the short range  proton-neutron correlations in the nuclear target~\cite{Amado:1976kn,Frankel:1976bq,Frankfurt:1977np,Frankfurt:1981mk,Frankfurt:1988nt,Frankfurt:2008zv}. These correlations lead to the presence of long tails in the nucleon momentum distributions   
inside a nucleus and thus extend the kinematic region of the emitted particles in p~+~$A$ reactions. It was also suggested that the cumulative effect is a result of 
the presence of multi-nucleon targets (``grains'') with masses $2m$, $3m$, $\ldots$ (where $m$ is the nucleon mass) inside the nucleus (see Refs.~\cite{Baldin:1978bw,Burov:1976xd,Efremov:1976xw,Efremov:1987mx, Braun:2001ru}). 
In both these explanations, all necessary requisites for the cumulative hadron production are present inside the nucleus before the p~+~$A$ reaction.

Our present study is based on an alternative model scenario~\cite{Gorenstein:1976zg}, see also Refs.~\cite{Gorenstein:1977hm, Bogatskaya:1978jh, Bogatskaya:1979qz, Anchishkin:1981fb,Golubyatnikova:1984yh, Kalinkin:1989wr,Motornenko:2016sfg,Motornenko:2016uzi}.  
The kinematically forbidden regions in $N+N$ collisions can be explored in p~+~$A$ reactions due to creation of heavy hadronic states and their successive collisions with nuclear nucleons. 
We assume that a heavy baryonic resonance $R$, created in a primary proton-nucleon (p~+~$N$) reaction, can propagate further through the nucleus, and it has a chance to interact with another nuclear nucleon earlier than its decay to a stable hadron occurs. Therefore, several nuclear nucleons are involved in the backward nucleon production.
Two mechanisms for the production of a backward nucleon $N(180\degree)$ will be considered: the reaction $R+N\rightarrow N(180\degree)+N$ and the resonance decay to nucleon and pion $R\rightarrow N(180\degree)+\pi$.

Excitations of heavy  baryonic states and their subsequent  re-scattering  can be also probed in the sub-threshold production~\cite{Aichelin:1986ss,Mosel:1992rb,Hartnack:1993bq} 
of massive hadrons in $A+A$ collisions.
Recently this approach was used to describe the data on strange and charm particle production~\cite{Steinheimer:2015sha,Steinheimer:2016jjk,Gallmeister:2017ths}.
The two phenomena  in p~+~$A$ reactions -- the production of hadrons outside the kinematic region of $N+N$ collisions and the sub-threshold production of strange and charmed heavy hadrons -- have probably the same origin, namely, the creation of heavy resonances and their further interactions with nuclear nucleons. 

This paper is organized as follows. In Sec.~\ref{sec:kin}
we calculate the maximal energy for the backward production of nucleons in p~+~$A$
reactions if $n=2,3,\ldots$ nuclear nucleons are involved in this reaction.
The mass of the baryonic resonance needed for this production is also calculated.
These results are obtained as a consequence of energy-momentum conservation.
In Sec.~\ref{sec:urqmd} the results of the Ultra-relativistic Quantum Molecular Dynamics simulations are presented.
Section \ref{sec:sum} summarizes the paper.

\section{Kinematic restrictions for the backward nucleons}
\label{sec:kin}

In this section we consider general restrictions on the energy of the backward nucleon emitted
at 180$\degree$ relative to the direction of the projectile proton in p~+~$A$ reactions in the rest frame of the target nucleus.
The restrictions obtained are a consequence of the energy-momentum conservation laws.
We are interested in the maximal value of the backward nucleons energy.
The production of any additional particles and/or a presence of a non-zero transverse particle momentum in the final state would
require extra energy and would cause a reduction of the final energy of the backward nucleon. Thus,
to find the maximal value of its energy we assume that no new hadrons are created,
and all nucleons move longitudinally.
Therefore, our kinematic analysis is reduced to the one-dimensional (longitudinal) problem.
Besides, we assume
$m = 0.94$~GeV for the nucleon mass and neglect the small difference between proton and neutron masses.

\subsection{$R+N\rightarrow N+N(180\degree)$}
The energy-momentum conservation in the reaction p~+~$N\rightarrow N +\ldots$ does not permit the backward nucleon production in the target nucleon rest frame.
If two nuclear nucleons are involved in the p~+~$A$
reaction the conservation laws for energy and momentum are:
\eq{
\label{en-1}
& \sqrt{p^2 + m^2} + m + m = \sqrt{k_2^2 + m^2} + \sqrt{p_1^2 + m^2} + \sqrt{p_2^2 + m^2}~,~~~~~
& p = p_1 + p_2 - k_2~,
}
where $p$ is the momentum of the projectile proton, $k_2>0$ is the momentum of the backward nucleon $N(180\degree)$, the lower index in $k_2$ denotes the number of nuclear nucleons involved,
whereas $p_1$ and $p_2$ are the final longitudinal momenta of the two nucleons.

Let us denote the maximal value of $k_2$ as $k_2^*$.
From Eq.~\eqref{en-1} and the maxima criteria $\partial k_2/ \partial p_1=\partial k_2/ \partial p_2=0$
one finds nucleon momenta $p_1 = p_2 = (p + k_2)/2$
which maximize the backward nucleon momentum $k_2=k_2^*$.
This leads to the following algebraic equation for the maximal kinetic energy,
$E_2^*\equiv \sqrt{(k^*_2)^2+m^2}-m$, of the backward nucleon:
\begin{equation}
E_2^* = m + \sqrt{p^2 + m^2} - 2~\sqrt{m^2 + \left(\frac{p + k_2^*}{2} \right)^2}~.
\label{E_max}
\end{equation}

\begin{figure}[h!]
\center
\includegraphics[trim={0 0 1.7cm 1.5cm},width=0.48\textwidth]{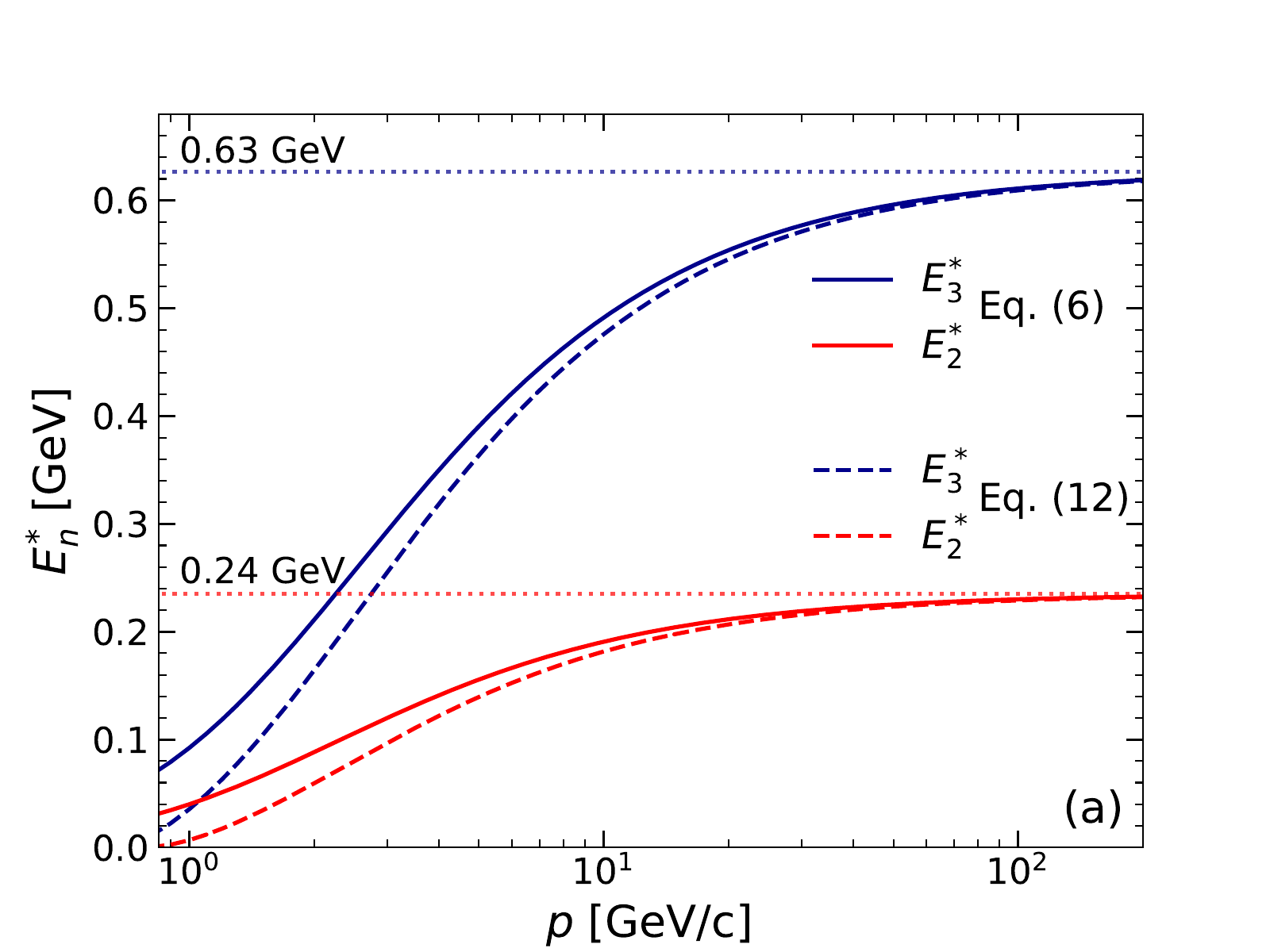}
\includegraphics[trim={0 0 1.7cm 1.5cm},width=0.48\textwidth]{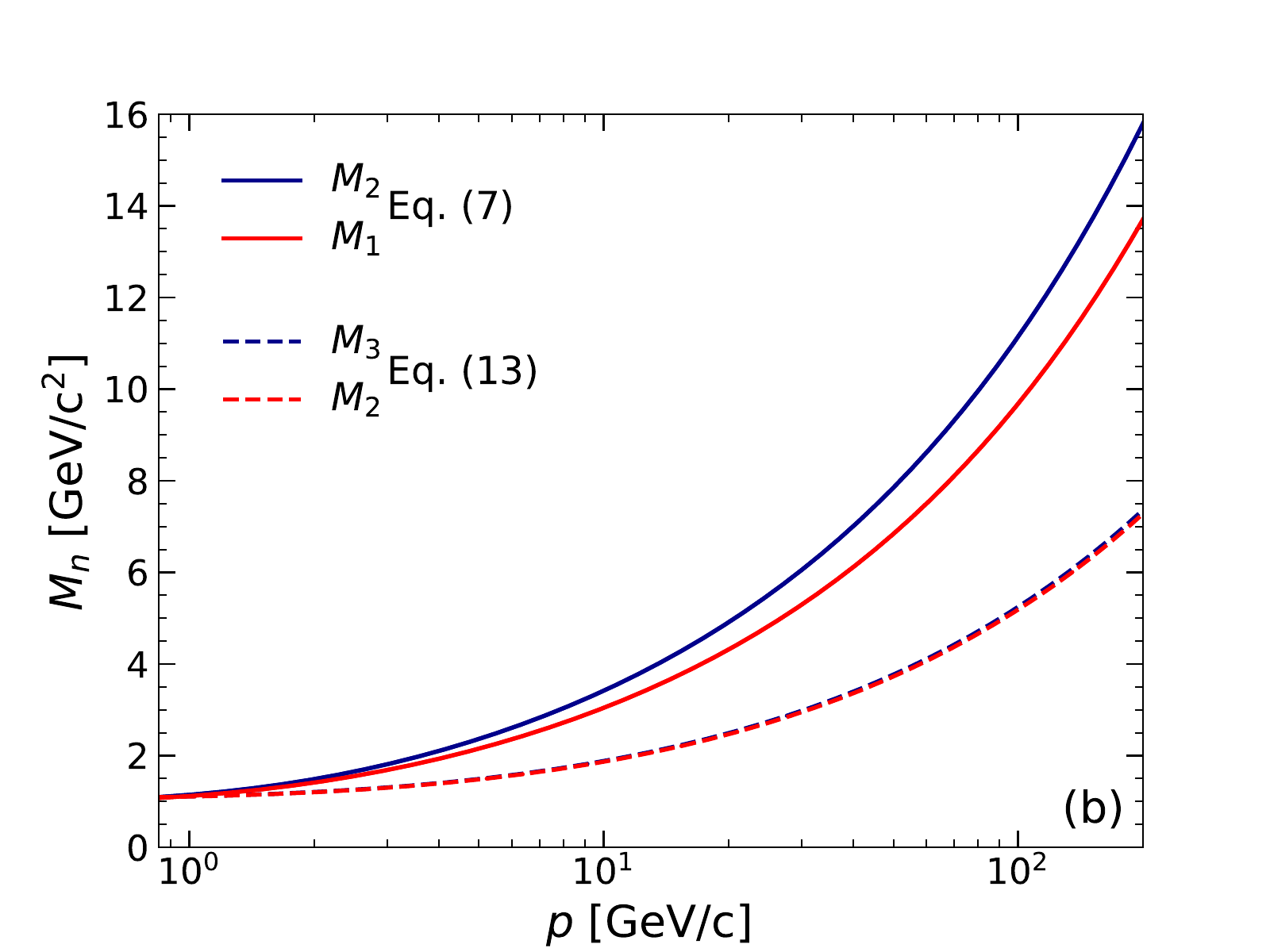}
\caption{(a) Maximal kinetic energies $E^*_2$ and $E_3^*$ of the backward nucleons given by Eq.~\eqref{En} are shown as functions of the projectile proton momentum $p$ by  solid red and blue lines, respectively. 
Horizontal dotted lines show the upper limits at $p \rightarrow \infty$. 
(b) The resonance masses $M_{1}$ and $M_2$ given by Eq.~\eqref{M-2} are shown by solid red and blue lines, respectively. 
Dashed lines on (a) and (b) represent the same quantities but given by Eqs.~\eqref{En_1} and \eqref{M-2_1}, when the additional production of a pion, $R\rightarrow N(180\degree) + \pi$, takes place.}
\label{fig-1}
\end{figure}

The solution of Eq.~\eqref{E_max} for the kinetic energy, $E^*_2$,  is presented in Fig.~\ref{fig-1} (a) by the lower
solid (red) line.

The value of $E_2^*$ increases with the projectile proton momentum $p$, and the
upper limit $E_2^* \cong 0.24$~GeV is reached at $p\rightarrow \infty$.

We assume that a backward nucleon with kinetic energy $E_2^*$ is created through a two-step process. First, the reaction p~+~$N\rightarrow R+N$
takes place, and a resonance $R$ with the mass $M_1$ is created. 
The backward nucleon production takes then place at the second step in the following reaction\footnote{ Note that the
reactions $\Delta +N\rightarrow N+N$  were discussed in Ref.~\cite{Gazdzicki:1997sg}
where these reactions were
proposed as a mechanism of the pion suppression in nucleus-nucleus reactions
at small collision energies  (see also  Ref.~\cite{Weyer:1990ye} ). }:
\eq{\label{R}
R+N~\rightarrow ~ N(180\degree) ~+~ N~.
}

To reach the maximal energy $E_2^*$ \eqref{E_max} of the backward nucleon the baryonic resonance mass $M_1$ after the 
first p~+~$N$ collision should be equal to:

\begin{equation}
M_1^2 = \left[\sqrt{p^2+m^2} - \sqrt{\left(\frac{p + k_2^*}{2}\right)^2 + m^2} + m\right]^2
- \left[p - \left(\frac{p + k_2^*}{2}\right)\right]^2~.
\label{M-1}
\end{equation}

The  resonance mass $M_1$ after the p~+~$N$ collision is shown in Fig.~\ref{fig-1} (b)  as a function of
the projectile momentum $p$
by the lower solid (red) line.

If $n\ge 2$ nucleons are involved in the reaction p~+~$A\rightarrow N(180\degree) +\ldots$  with a backward nucleon production,
the energy and momentum conservation equations are:
\begin{equation}
\sqrt{p^2 + m^2} + n\, m = \sqrt{k_n^2 + m^2} + \sum_{i = 1}^{n} \sqrt{p_i^2 + m^2}~,~~~~~
p =  \sum_{i = 1}^{n} p_i - k_n~. \label{enr}
\end{equation}

Similar to the case of $n=2$, one finds the nucleon momenta $p_1 = p_2=\ldots = p_n = (p + k_n)/n$ which maximize the backward
nucleon momentum $k_n=k_n^*$.
This leads to the following equation for
the maximal kinetic energy  of the backward nucleon:

\eq{
E_n^* = (n-1)\, m + \sqrt{p^2 + m^2} - \,\sqrt{n^2m^2 + \left({p + k_n^*} \right)^2}~.
\label{En}
}

A solution of this equation for $n=3$ is presented in Fig.~\ref{fig-1} (a) by an upper solid (blue) line.
The value of $E_3^*$ increases with the projectile proton momentum $p$, and the
upper limit $E_3^* \cong 0.63$~GeV is reached at $p\rightarrow \infty$.
Note also that the resulting conservation laws and final expression ~\eqref{En} for $E_n^*$ are the same
as for a collision of the projectile proton with $n$-nucleon "grain" in a nucleus.

To produce a backward nucleon with energy  $E_n^*$  (\ref{En}) in reaction \eqref{R} a baryonic resonance with mass $M_{n-1}$ needs to be formed in $n-1$ preceding collisions with nuclear nucleons.
For example, to reach the energy $E_3^*$ within the reaction \eqref{R}  
the following two preceding reactions have to take place: 
p~+~$N\rightarrow R_1+N$
and then $R_1+N\rightarrow R_2+N$.

The straightforward calculations give:

\eq{M_{n-1}^2 & ~=~  \left[\sqrt{p^2+m^2} -(n-1)\left( \sqrt{\left(\frac{p + k_n^*}{n}\right)^2 + m^2} - m \right) \right]^2 \nonumber \\
& ~ - ~  
\left[p - (n-1)\left(\frac{p + k_n^*}{n}\right)\right]^2~.
\label{M-2}
}

For $n=2$, Eqs.~\eqref{En}  and \eqref{M-2} are reduced to Eqs.~\eqref{E_max} and \eqref{M-1}, respectively. 
The  resonance mass $M_2$ after two successive  collisions with nuclear nucleons
is shown in Fig.~\ref{fig-1} (b)  as a function of the projectile proton momentum $p$
by the upper solid (blue) line.

From Fig.~\ref{fig-1} one observes that $E_n^*$  increases strongly with the projectile
proton momentum $p$ up to $p\sim 10$~GeV/$c$. This corresponds to a mass region of the
baryonic resonances not larger than $(3-4)$~GeV. A further strong increase of the baryonic resonance masses
at $p> 10$~GeV/$c$ seen in Fig.~\ref{fig-1} (b) leads to only a slight increase of $E_n^*$ shown
in Fig.~\ref{fig-1} (a).

\begin{figure}[h!]
\center
\includegraphics[trim={0 0 1.7cm 1.5cm},width=0.6\textwidth]{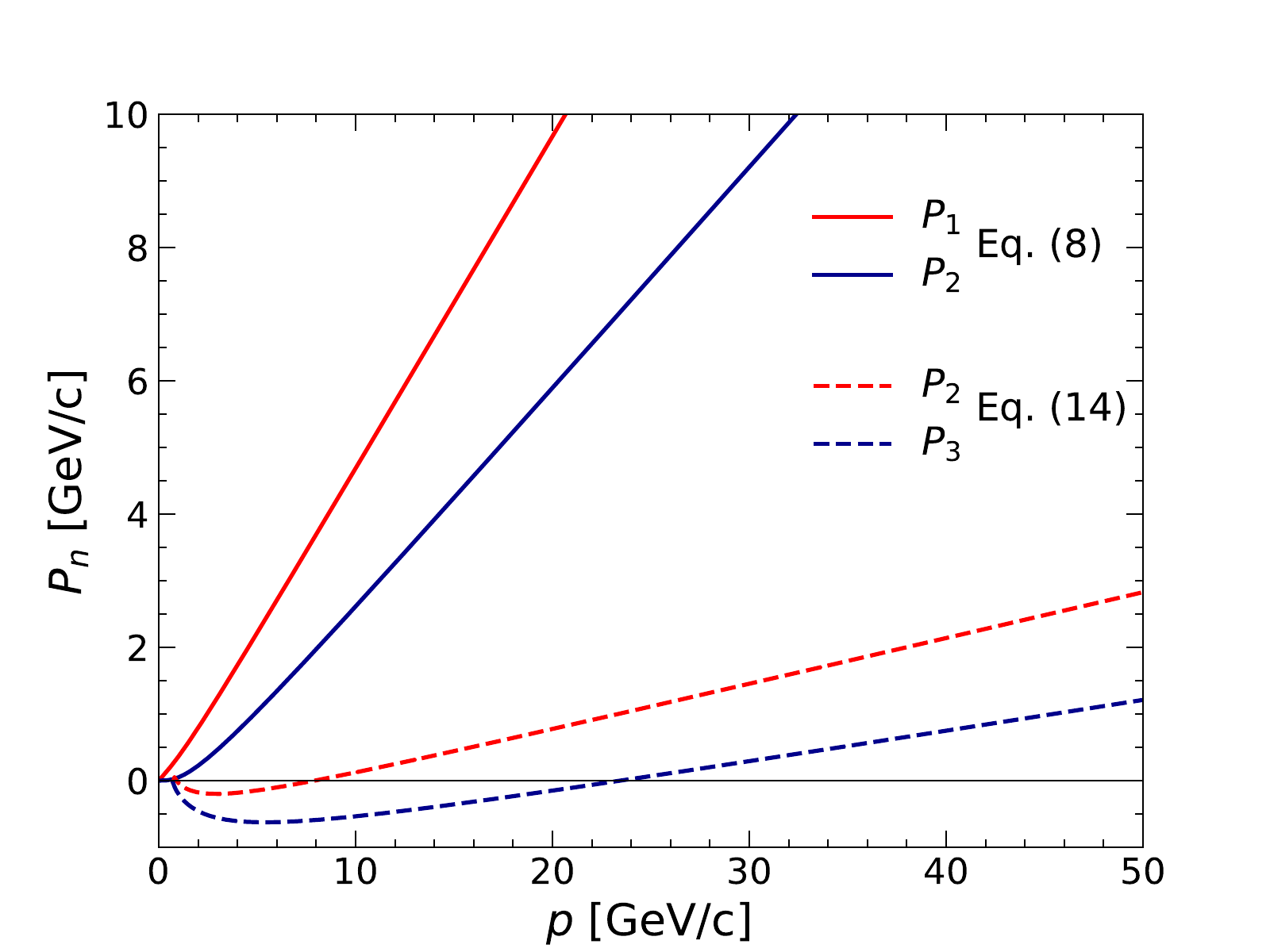}
\caption{Solid lines present the momentum $P_{n}$~\eqref{P-1} of baryonic resonance in reaction $R+N\rightarrow N+N(180\degree)$ for $n=2$ (upper solid red line) and $n=3$ (lower solid blue line).  Dashed lines present $P_{n}$~\eqref{P-2} in reaction $R\rightarrow N(180\degree) + \pi$ for $n=2$ (upper dashed red line) and $n=3$ (lower dashed blue line).}
\label{fig-mom}
\end{figure}

The momentum of a baryonic resonance  with mass $M_{n-1}$ after $n-1$ collisions with nuclear nucleons should be equal to:
\begin{equation}
P_{n-1} = p - (n-1)\,p_n = p - (n-1)~ \frac{p + k_n^*}{n}~.
\label{P-1}
\end{equation}
The solutions of Eq.~\eqref{P-1} for $n=2$ and $3$ are presented in Fig.~\ref{fig-mom}
as functions of the projectile proton momentum by the solid 
upper red and lower blue line, respectively.

\subsection{$R \rightarrow N(180\degree) + \pi$}

Let us assume now a resonance decay into the backward nucleon and pion.
If $n$ nuclear nucleons are involved, the conservation laws for the energy and momentum are:
\begin{equation}
\label{enr_1}
\sqrt{p^2 + m^2} + n\, m = \sqrt{k_n^2 + m^2} + \sum_{i = 1}^{n} \sqrt{p_i^2 + m^2}+ 
\sqrt{p_\pi^2 + m_\pi^2}~,
\end{equation}
\begin{equation}
p =  \sum_{i = 1}^{n} p_i + p_\pi - k_n~,
\label{enr_m_1}
\end{equation}
where $m_\pi$ and $p_\pi>0 $ are the pion mass and longitudinal momentum, respectively. 
At a given value of the projectile proton momentum $p$, the maximal value of the backward nucleon $k_n$ is reached
at the conditions:
$p_1 = p_2 = ...  = p_n,$ and $p_\pi/m_\pi = p_n/m~$,
i.e., the $n$ nuclear nucleons and created pion  should move with the same velocity.
The maximal kinetic energy $E_n^*$ of the backward nucleon $N(180\degree)$ and  the mass $M_n$ of a resonance before 
its decay into the backward nucleon and pion,
\eq{\label{dec} 
R\rightarrow N(180\degree) +\pi~,
}
are calculated as:
\begin{eqnarray}
 E_n^{*} & =&  (n-1) m + \sqrt{p^2 + m^2} - \sqrt{(p+k_n^*)^2 + \left(n m+m_\pi\right)^2}~,
\label{En_1} \\
 M_{n}^2 & =& \left[\sqrt{p^2+m^2} -n\left( \sqrt{\left(\frac{p + k_n^*}{n + m_\pi/m}\right)^2 + m^2} - m \right) \right]^2 \nonumber \\
 & - &
\left[p - n\left(\frac{p + k_n^*}{n + m_\pi/m}\right)\right]^2 ~.
\label{M-2_1}
\end{eqnarray}

The solutions of Eqs.~\eqref{En_1} and \eqref{M-2_1} for $n=2$ and $n=3$ 
as functions of the projectile proton momentum are presented in Figs.~\ref{fig-1} (a) and (b) by lower (red) and upper (blue) dashed lines, respectively.

The momentum of the baryonic resonance before its decay to the backward nucleons and pions is:
\begin{equation}
P_{n} = p - n\,p_n = p - n~\frac{p + k_n^*}{n+ (m_\pi/m)}~.
\label{P-2}
\end{equation}
The solutions of Eq.~\eqref{P-2} for $n=2$ and $n=3$ are presented in Fig.~\ref{fig-mom} by upper red and lower blue dashed lines, respectively.
One can see that the resonance should move backwards at small $p$ to produce a backward nucleon with maximal energy.

\section{UrQMD simulations of ${\rm p}+A$ reactions}
\label{sec:urqmd}
In this section an analysis of the backward production of protons within 
the Ultrarelativistic Quantum Molecular Dynamics  
(UrQMD) transport model~\cite{Bass:1998ca,Bleicher:1999xi} is performed. The mechanisms of the backward nucleon production by heavy baryonic resonances suggested in previous sections are probed by microscopical simulations of ${\rm p}+A$ collisions.
The UrQMD model performs simulations by the Monte-Carlo calculations of stochastic two-particle collisions and resonance decays, it includes nucleon $N^*$ and $\Delta$ resonances with pole masses in the region $1.232-2.250$~GeV, higher masses are possible due to the finite widths in Breit-Wigner distribution for unstable particles. The hadron-like states with masses higher than $ 2.250$ GeV are effectively modeled by the string excitation, this mechanism is dominating for inelastic hadron collisions at the center of mass energy of nucleon pair $\sqrt{s_{\rm NN}}>3$ GeV. Contrary to hadronic degrees of freedom, the string degrees of freedom do not interact with other objects, they are only subject to fragmentation \footnote{Note a novel string model for high energy p+A and A+A collisions where the string-string interactions are considered~\cite{Bierlich:2018xfw}.}. Therefore, in UrQMD simulations the  strings are not able to contribute to suggested mechanism of successive collisions with nuclear nucleons. Within UrQMD the Fermi motion of nucleons is modeled with the random distribution of nuclear nucleons momenta in the range of 0~-~300~MeV in a nucleus rest frame. The Fermi motion allows to widen the available kinematic region for backward proton production. An another theorized source of cumulative particle production, nuclear short range correlations~\cite{Frankfurt:1988nt}, cannot be studied within UrQMD because of a lack of implementation of the phenomenon. The implementation of the nuclear short range correlations in a transport model is still a conceptual problem and a subject for future studies.

\begin{figure}[h!] 
\centering
\includegraphics[trim={0 0 1.25cm 0},width=0.49\textwidth]{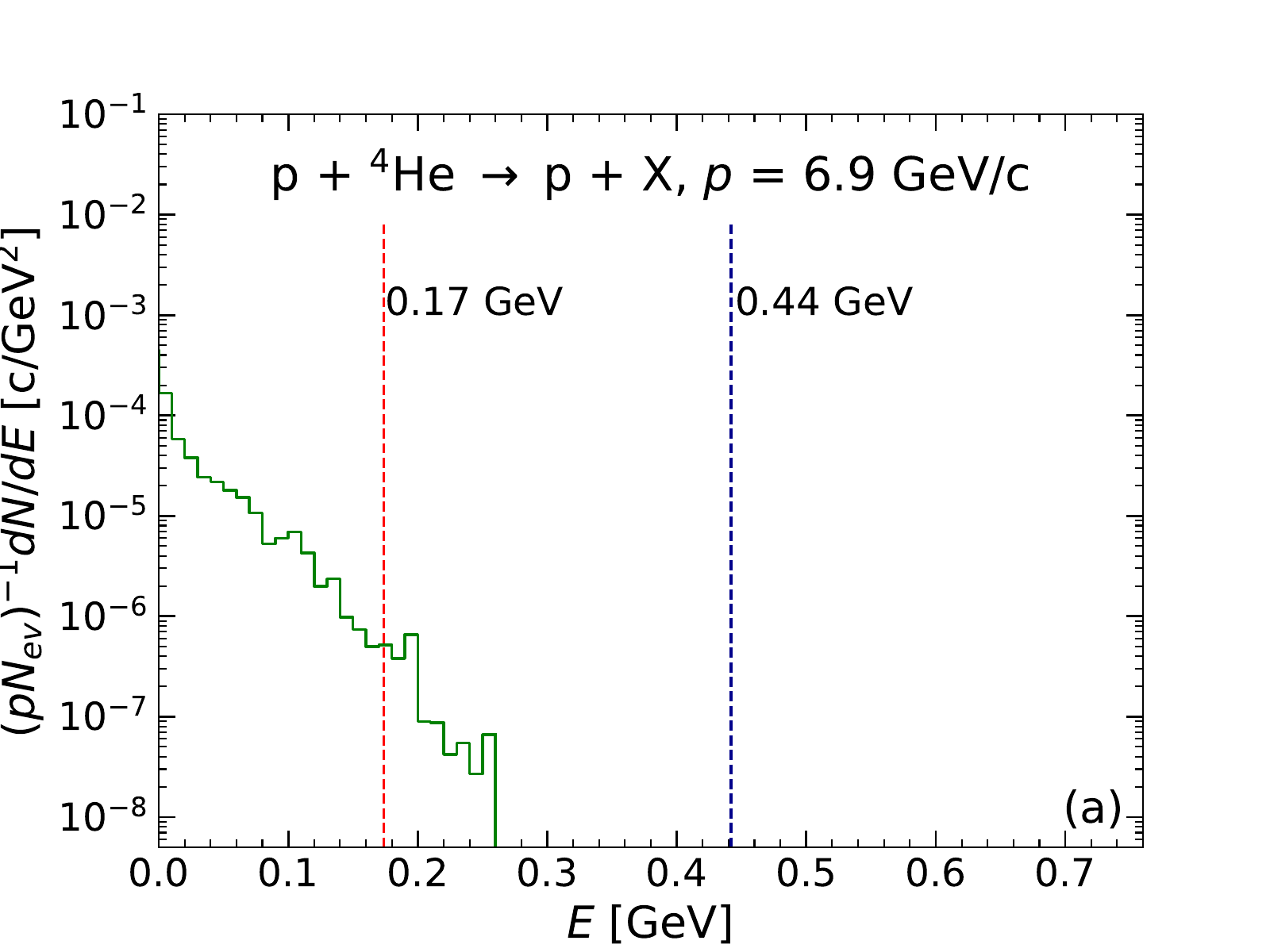}
\includegraphics[trim={0 0 1.25cm 0},width=0.49\textwidth]{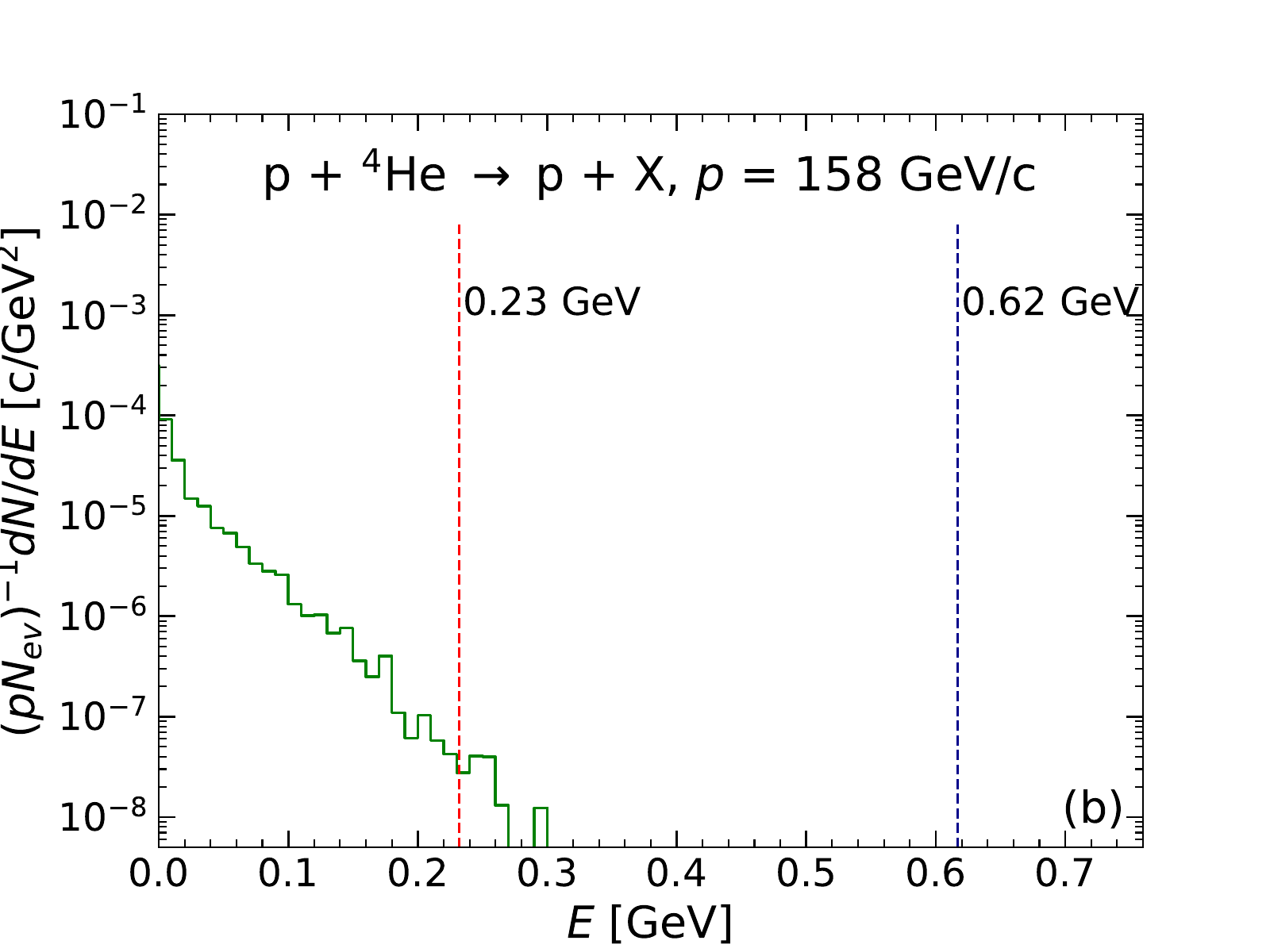}\\
\includegraphics[trim={0 0 1.25cm 0},width=0.49\textwidth]{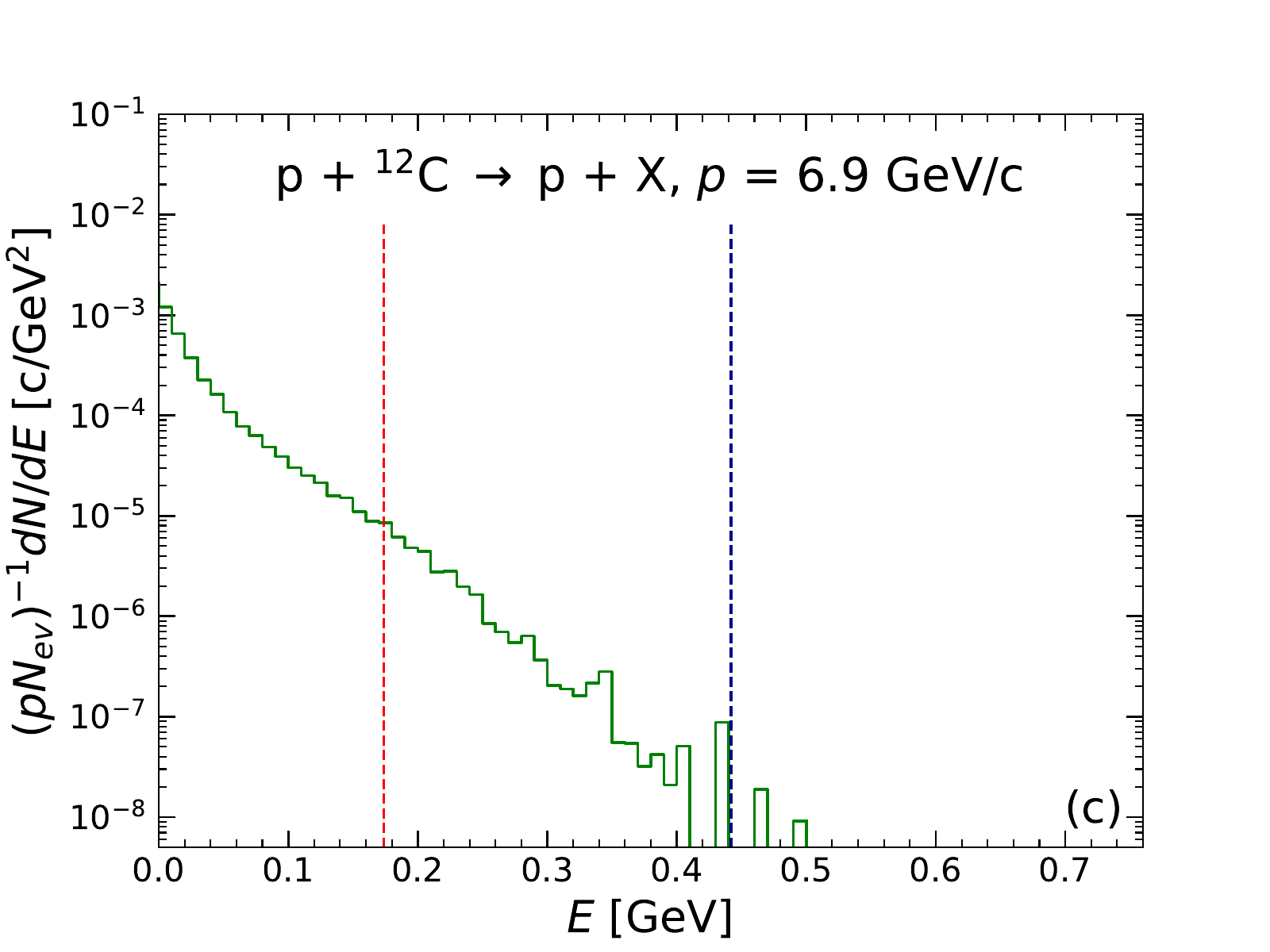}
\includegraphics[trim={0 0 1.25cm 0},width=0.49\textwidth]{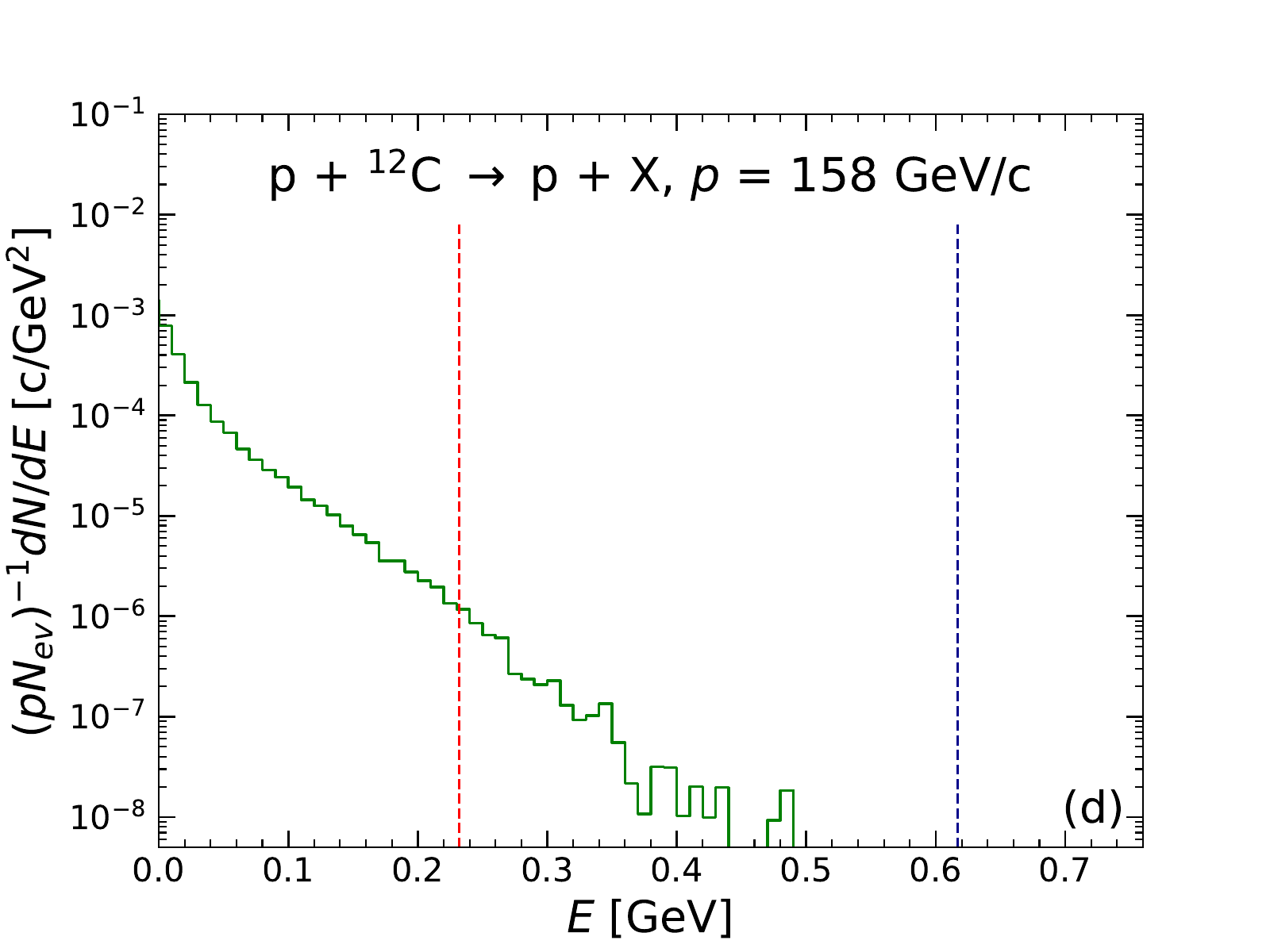}\\
\includegraphics[trim={0 0 1.25cm 0},width=0.49\textwidth]{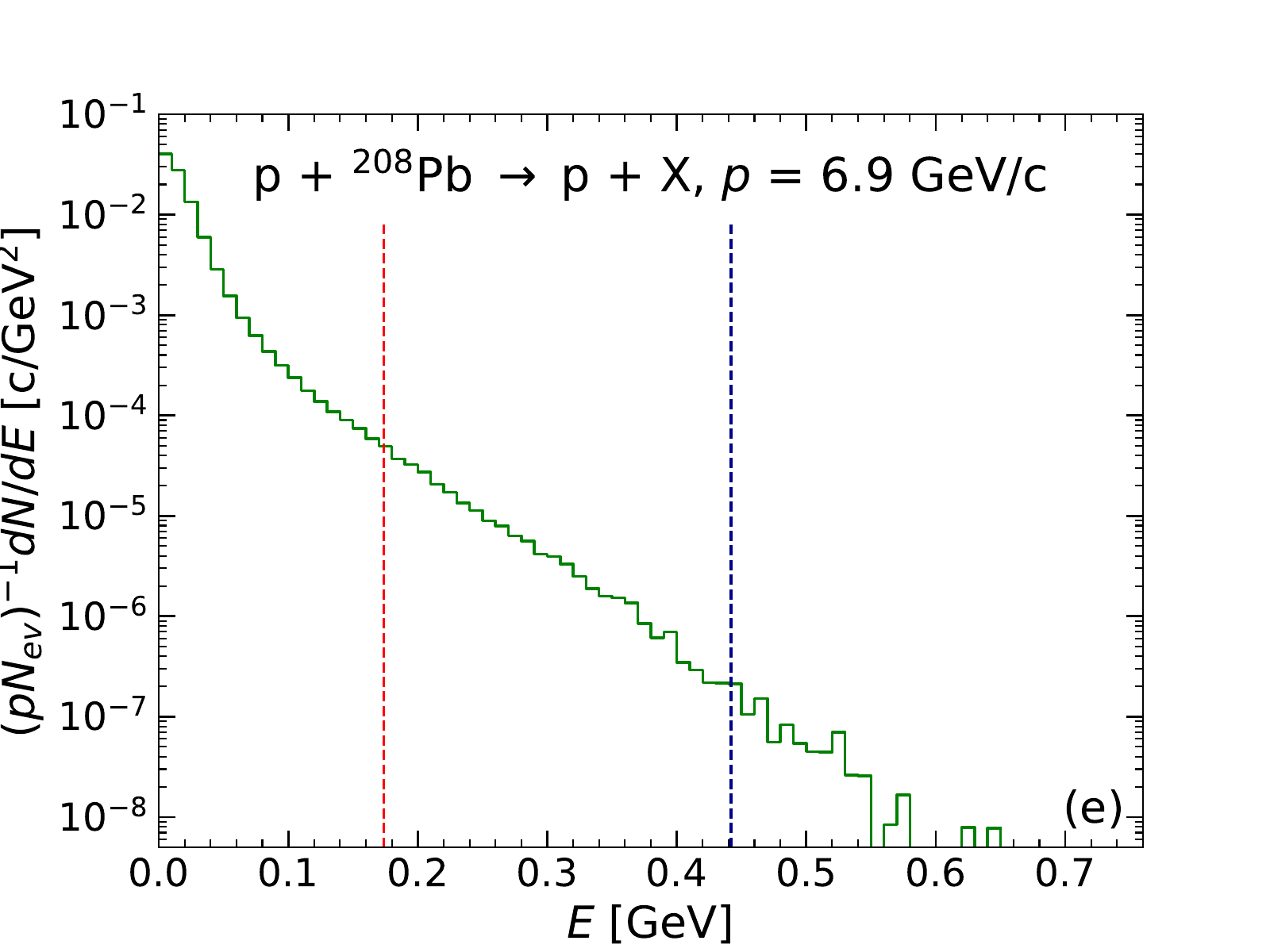}
\includegraphics[trim={0 0 1.25cm 0},width=0.49\textwidth]{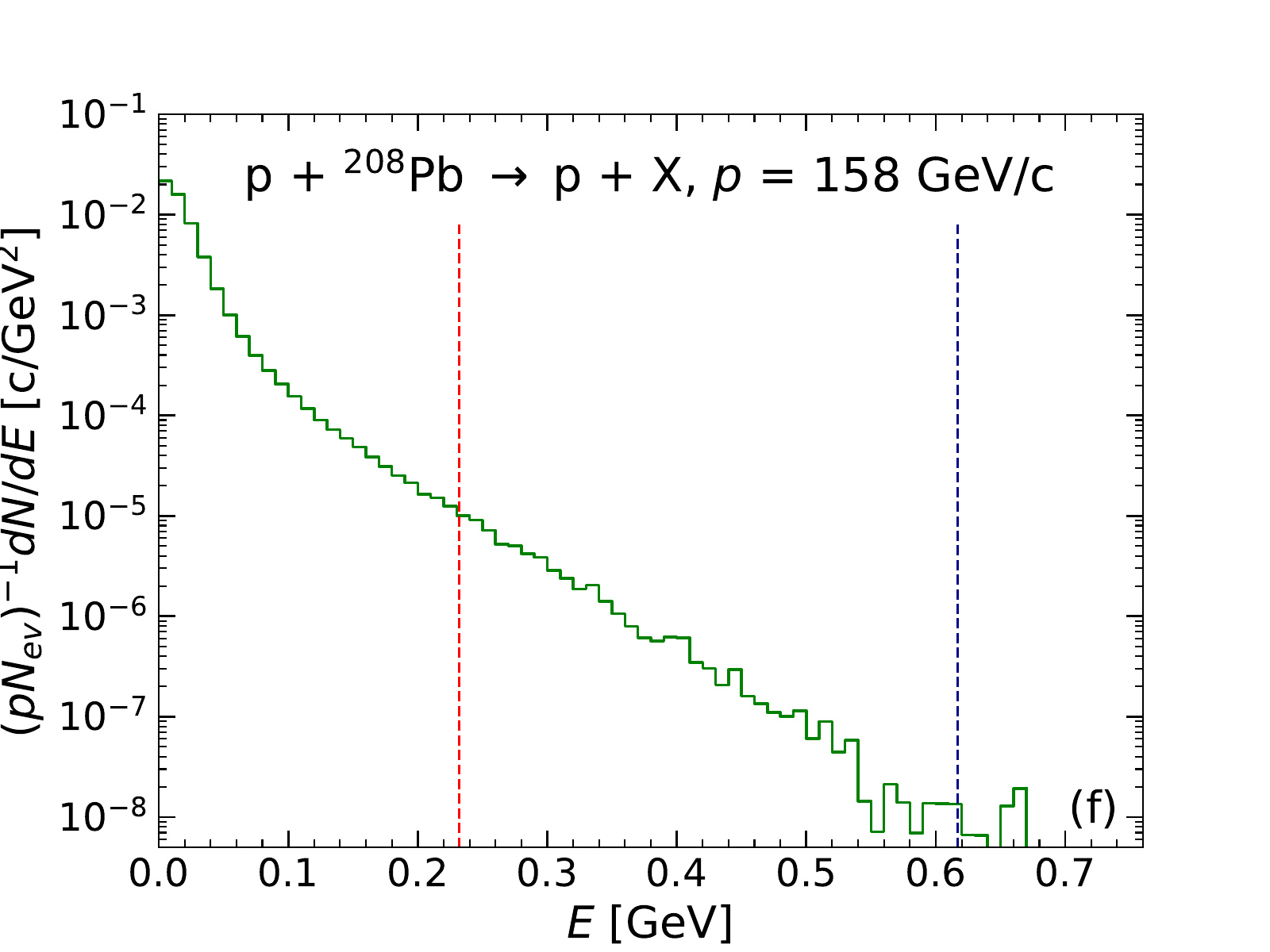}
\caption{Spectra of the backward protons as functions of their kinetic energy.
For each reaction a sample of $N_{ev} = 10^8$ central collision events with zero impact parameter is used.
The vertical dashed lines show the maximum kinetic energies of the backward
nucleon at the corresponding initial projectile proton momenta $p$ as calculated with Eq.~\eqref{En}, for collision numbers $n=2$ (red lines) and $n=3$ (blue lines).
}
\label{fig-UrQMD}
\end{figure}

For the simulations of ${\rm p}+A$ energetic collisions the standard setup of UrQMD-3.4 is used. In this setup the mean fields and the hydrodynamic stage of collision are not considered. The backward nucleons are defined as nucleons emitted in a narrow cone with respect to beam axis, $180^{\degree}\pm 6^{\degree}$. 
The UrQMD simulations are performed in ${\rm p+A}$  collisions with different nuclei ${\rm ^4He}$, ${\rm ^{12}C}$, and ${\rm ^{208}Pb}$  at the two values of the projectile proton momenta $p=6.9$~GeV/$c$ and $p=158$~GeV/$c$. These two momenta correspond to available energies at Dubna Synchrophasotron and the top CERN SPS energy, respectively. 
The backward nucleon production is a rare phenomenon and requires a large sample of events to study energetic cumulative particles. For each of p~+~$A$ reaction a number of $N_{\rm ev}=10^8$ collision events was simulated. To enlarge a production of the backward nucleons the  central p~+~$A$ collisions with zero impact parameter $b=0$ are mostly considered.
The calculated spectra are presented in Fig.~\ref{fig-UrQMD}. To avoid experimental problems with neutrons,  only the backward  protons are considered in the analysis. Therefore, a straightforward comparison with experimental data can be done. 

The UrQMD results presented in Fig.~\ref{fig-UrQMD} suggest that the  backward proton  spectra increase strongly with the atomic number $A$ of the target nucleus. Both the number of the backward protons and their largest kinetic energy increase with $A$. This behavior goes in line with the results from the analysis in Sec.~\ref{sec:kin} as  the possible number of primary and successive collisions with nuclear nucleons increases strongly with $A$. 
The maximal energy of the backward protons does not show a noticeable increase with $p$ from $p$~=~6.9~GeV/$c$ to $p$~=~158~GeV/$c$. This observation just reflects an absence of baryonic resonances with a mass larger than $M>3$~GeV in the present version of the UrQMD model.
An inclusion of resonances with masses larger than $M>3$~GeV or an implementation of string-hadron interactions in the UrQMD simulatioms  would allow to widen the kinematic range for cumulative particles. This is a point for future studies. 

\begin{figure}[h!]
\center
\includegraphics[width=0.496\textwidth]{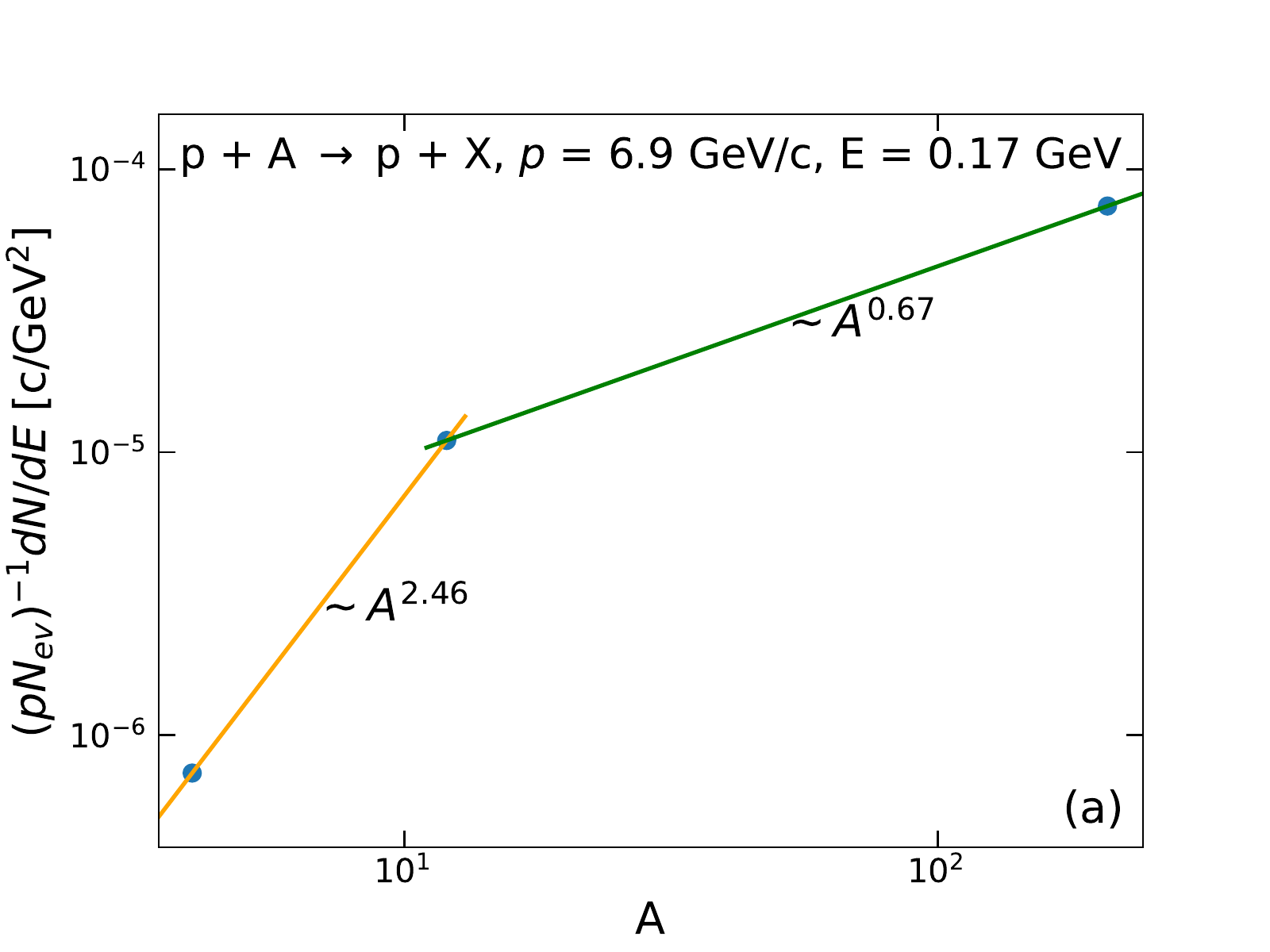}
\includegraphics[width=0.496\textwidth]{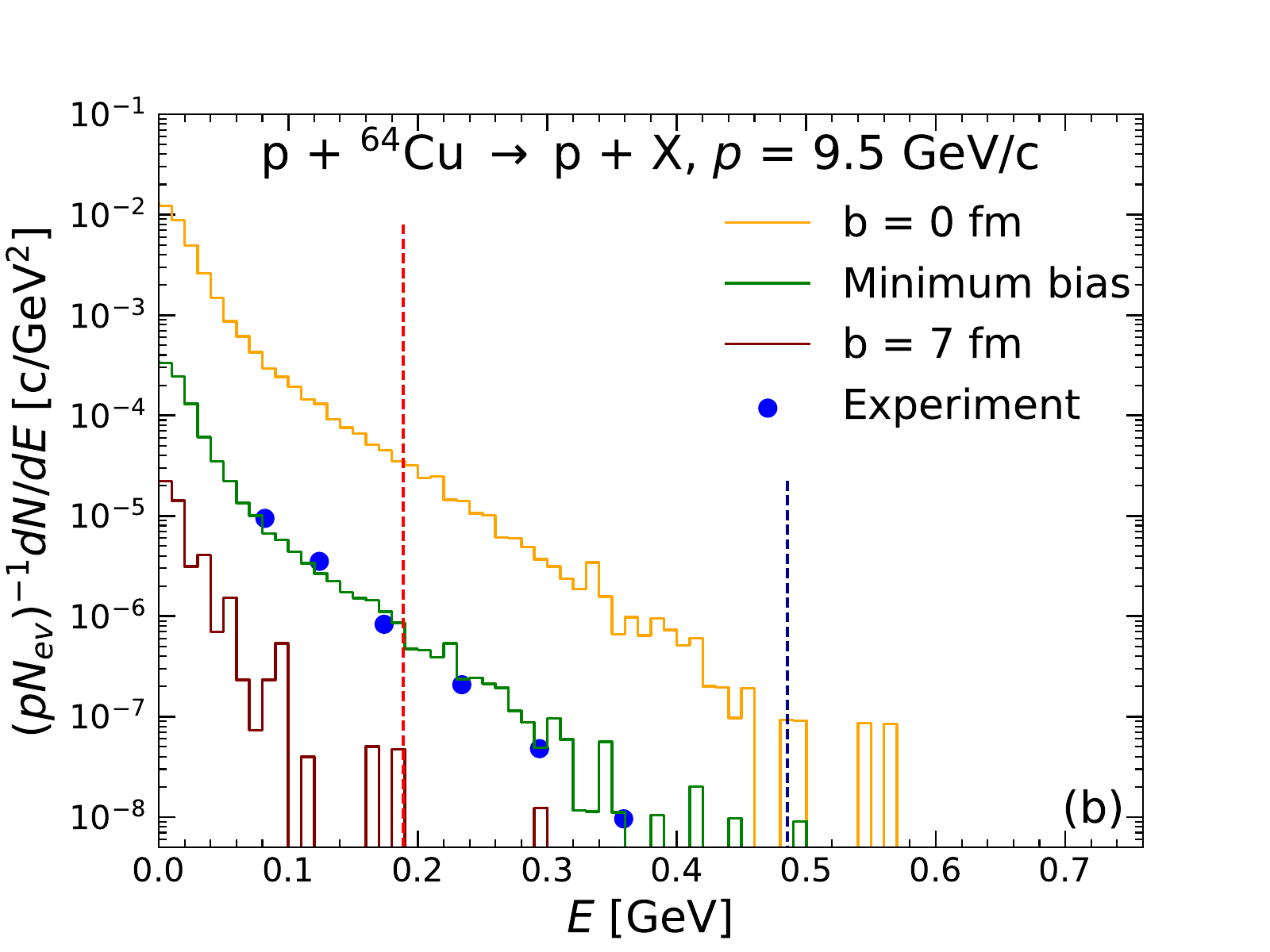}
\caption{ (a) The backward proton spectra  at the fixed kinetic energy $E=E_2^*=0.17$~GeV in p~+~He, p~+~C, and p~+~Pb collisions at $p=6.9$~GeV/$c$. (b) Comparison of the UrQMD results for proton spectra at $180\degree$ with the data   \cite{Frankel:1979uq} in  p~+~Cu reactions at $p =  9.5$ GeV/$c$. The histograms
correspond to the UrQMD results for peripheral ($b=7$~fm), minimum bias, and central ($b=0$), from below to up. The vertical dashed lines show the values of $E=E_2^* = 0.19$ GeV and $E= E_3^* = 0.49$ GeV
calculated with Eq.~(\ref{En}). }
\label{exp-UrQMD}
\end{figure}

In Fig.~\ref{exp-UrQMD} (a) the UrQMD values of the backward proton spectra  at the kinetic energy $E=E_2^*=0.17$~GeV are presented in p~+~$A$ collisions at $p=6.9$~GeV/$c$ as a function of $A$. Approximating these values by  $\sim A^\alpha$ dependence, one finds $\alpha\cong  2.46$ for light nuclei (from He to C) and $\alpha\cong 0.67$ for heavy nuclei (from C to Pb). 

The UrQMD results demonstrate also a strong dependence of the backward proton spectra on the centrality in p~+~$A$ reactions.  In Fig.~\ref{exp-UrQMD} (b) we present the backward proton spectra for UrQMD simulations in p~+~Cu reactions at the projectile momentum $p=9.5$~GeV/$c$. The lower (red) histogram presents the UrQMD results for the peripheral collisions with impact parameter $b=7$~fm, the upper (yellow) histogram for the central collisions with $b=0$, while the intermediate (green) histogram for the minimum bias p~+~Cu reactions. These minimum bias UrQMD results are in a good agreement with the data 
presented in Ref.~\cite{Frankel:1979uq} and shown in Fig.~\ref{exp-UrQMD} as full circles.
Note, however, that in our calculations a backward proton is defined as observed at angles of
$180\degree \pm 6\degree$.  
We have to use an additional normalization factor for the measured data to compare them with our results.

\begin{figure}[h!]
\center
 \includegraphics[width=0.496\textwidth]{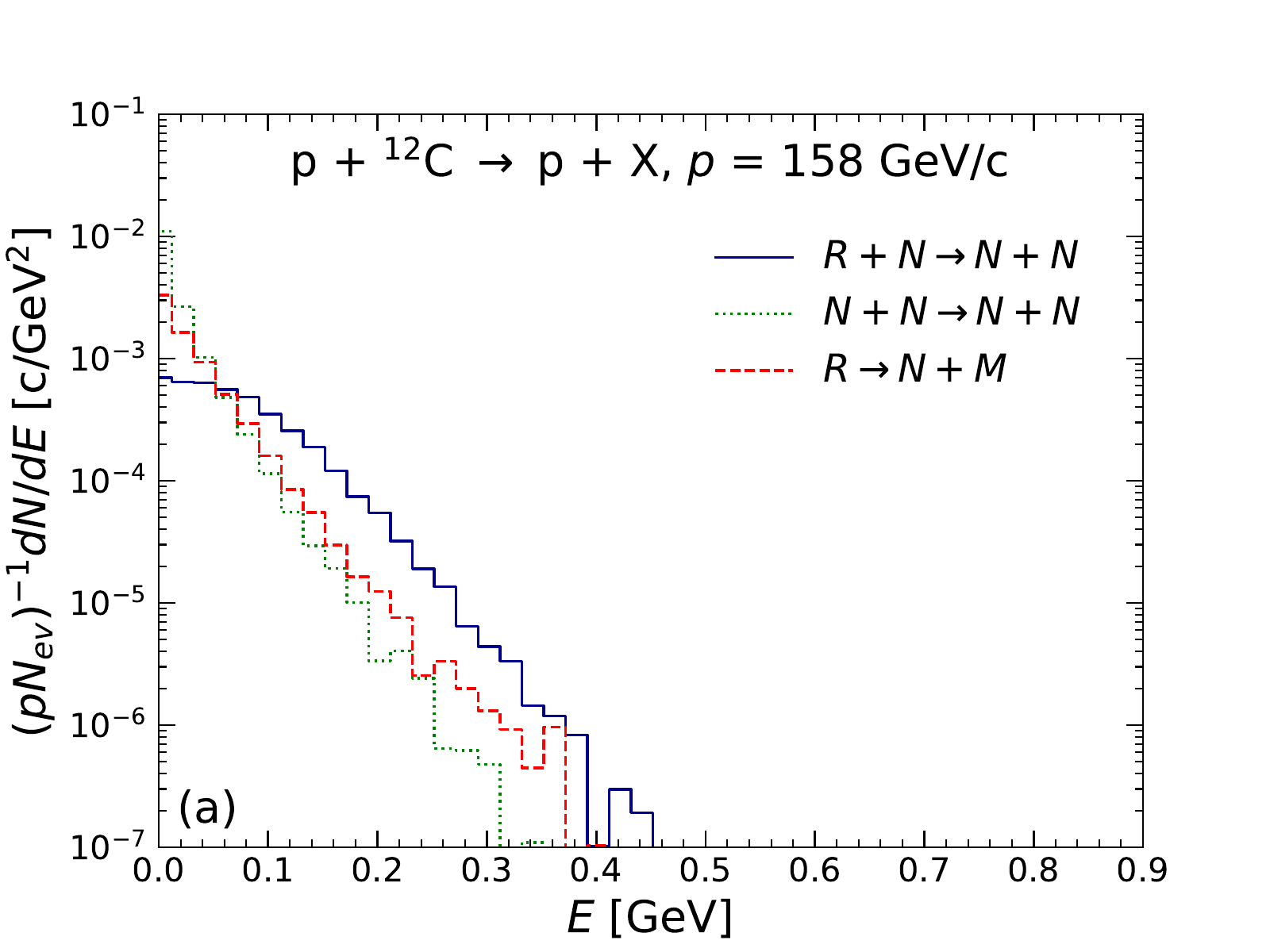}
 \includegraphics[width=0.496\textwidth]{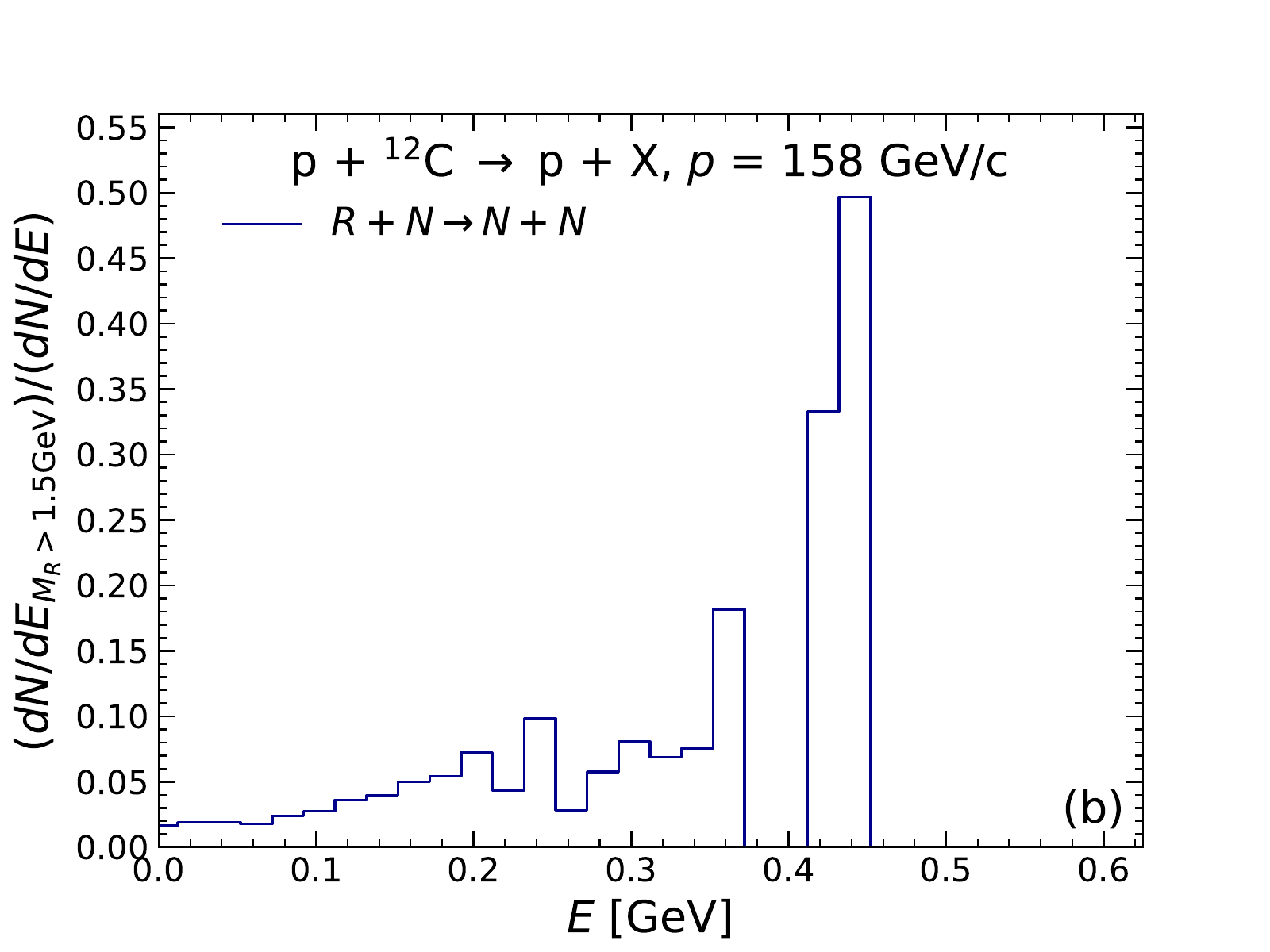}
 \includegraphics[width=0.496\textwidth]{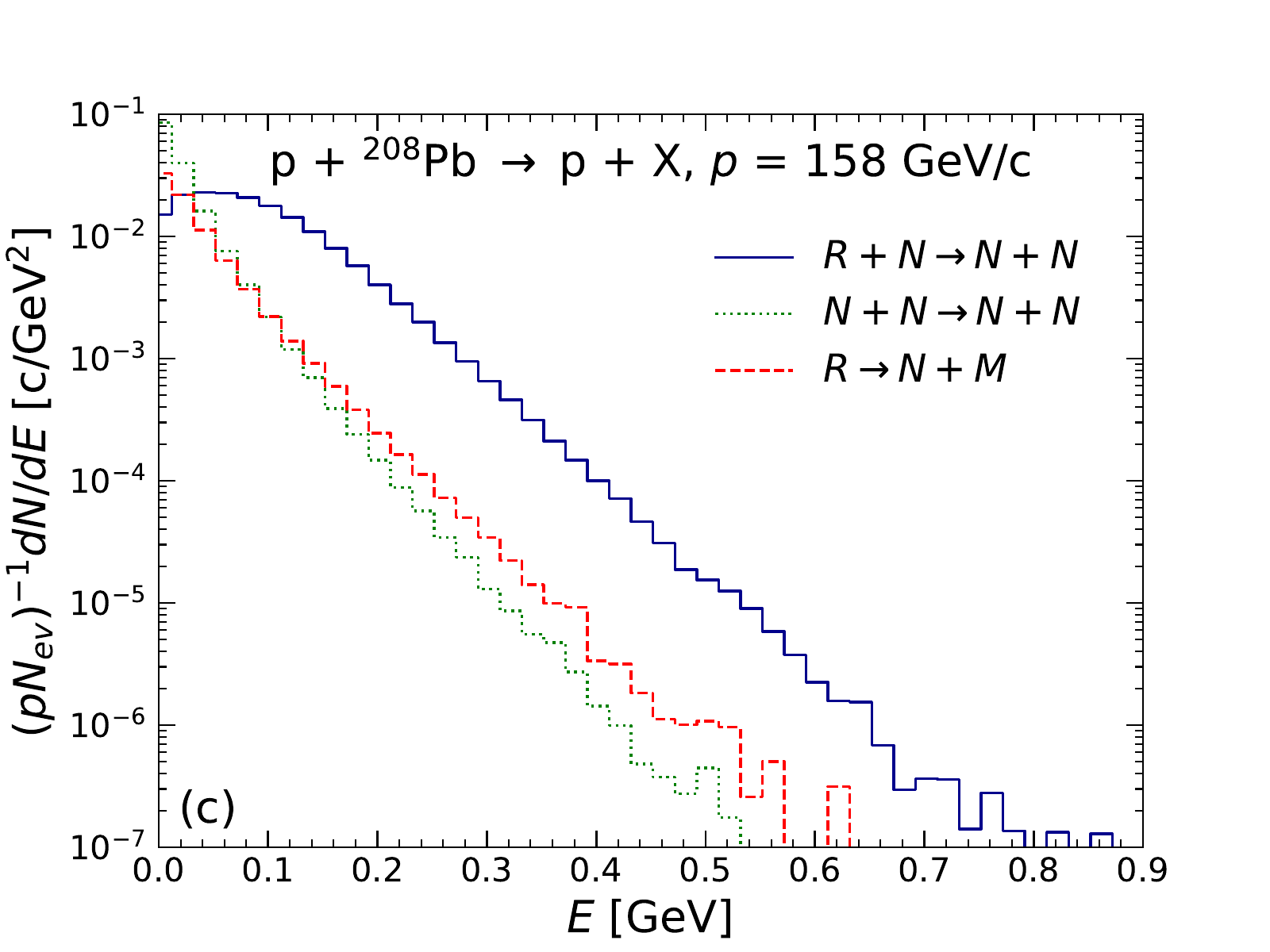}
 \includegraphics[width=0.496\textwidth]{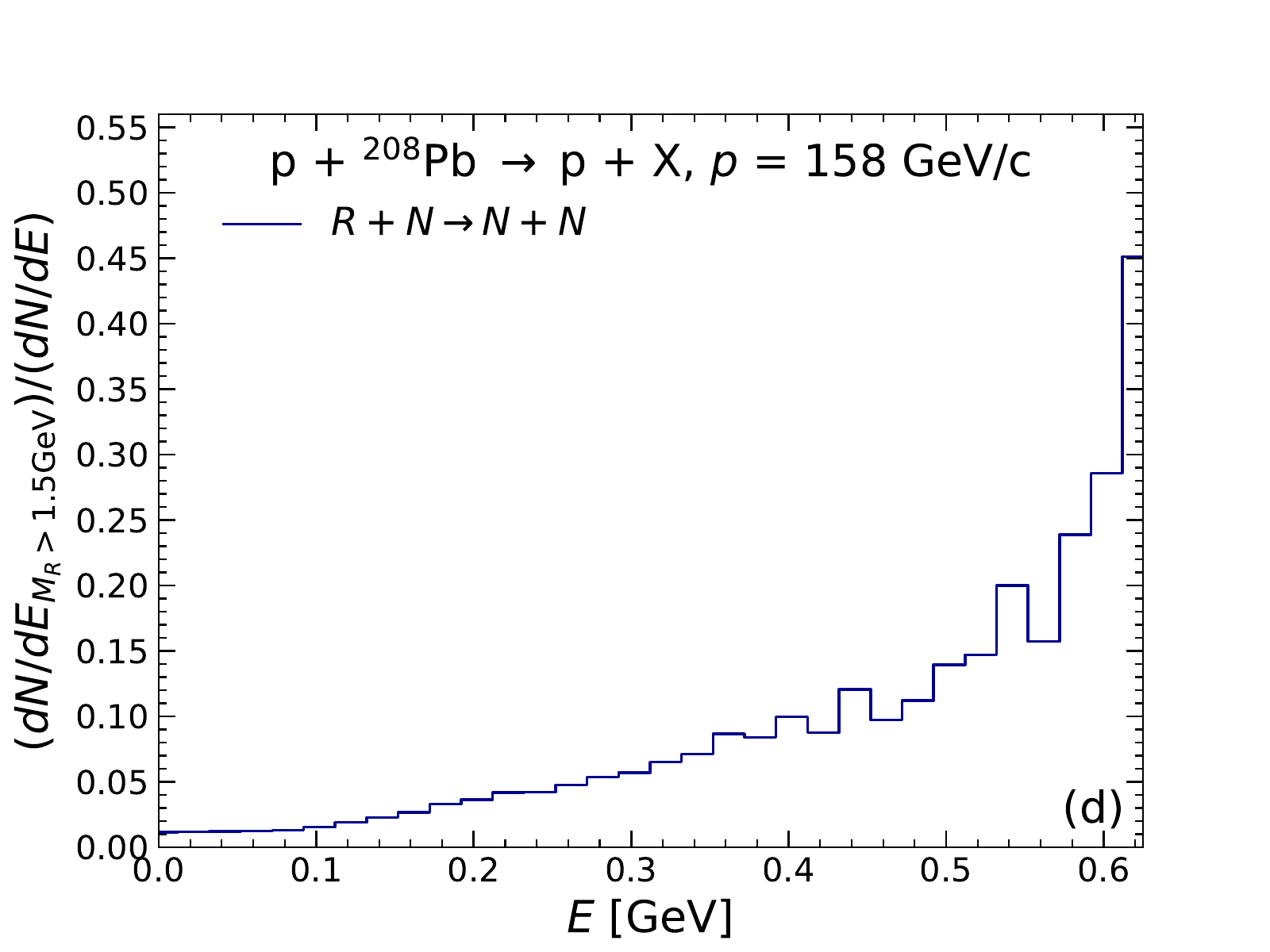}
 \caption{(a) and (c) The UrQMD spectra of the backward protons ($180^{\degree}\pm 15^{\degree}$) produced from the different sources as functions of the backward proton kinetic energy in p~+~C (a) and p~+~Pb (c) collisions at the projectile momentum $p=158$~GeV/$c$.
 (b) and (d) present the ratios of the spectra of the backward  protons  created in reactions $R+N\rightarrow N+N$ by heavy resonances $R$ with masses $M>1.5$~GeV  to the spectra of all backward protons created in this type of reactions.}
\label{fig:sources}
\end{figure}
Figures~\ref{fig:sources} (a) and (c) show the  contributions of different sources to the final spectrum of the backward protons as functions of the proton kinetic energy at the projectile proton momentum 158~GeV/$c$ in p~+~C and p~+~Pb reactions, respectively. 
To enlarge the event statistics we define now backward protons as being emitted within the cone of backward angles $180^{\degree}\pm 15^{\degree}$.

The Fermi motion of nucleons inside nuclei  with momenta up to 
$\sim 300$~MeV is implemented in the UrQMD model. This allows to produce backward nucleons even in a primary p~+~$N$ collision with the nuclear nucleons. On the other hand, the strong nucleon motion inside the nucleus due to the short range correlation effects are absent in the present version of the UrQMD model. The multi-nucleon target `grains' inside nucleus  are also absent. 

From Figs.~\ref{fig:sources} (a) and (c) one observes that there are three sources of the backward protons:  (multiple) elastic $N+N$  (re)scatterings which take into account the  Fermi motion of nuclear nucleons, the resonance decays $R\rightarrow N+\ldots$, and the reactions  $R+N\rightarrow N+N$.
At low kinetic energies of the backward proton,  elastic (re)scattering effects dominate. On the other hand, at larger $E$-values the  backward proton production due to reactions $R+N\rightarrow N+N$ becomes the dominant source.  

Figures ~\ref{fig:sources} (b) and (d) show  the ratios of the spectra of the backward  protons created in reaction $R+N\rightarrow N+N$ by resonances with heavy masses $M>1.5$~GeV to the backward protons produced in this type of reactions by all resonances $R$ with any mass. A partial contribution from heavy resonances with $m>1.5$~GeV appears to be rather small, but it strongly increases with $E$.

\section{Summary}
\label{sec:sum}

 The production of backward nucleons $N(180\degree)$ in p~+~$A$ collisions in the nuclear target rest frame  is studied. This kinematic region is forbidden in 
 p~+~$N$ reactions. It is suggested that the backward nucleons can be created by heavy baryonic resonances produced in several successive collisions with nuclear nucleons. A baryonic resonance formed in a primary p~+~$N$ collision propagates through the nucleus and can interact with other nuclear nucleons earlier than it decays. Thus, several nuclear nucleons, $n=2,3,\ldots $, are involved in the backward nucleon production.  To find the largest possible energy of the backward nucleons in such a scenario the two competitive mechanisms are considered. The first one assumes that the baryonic resonance $R$ is formed in the $n-1$ preceding collisions, and then it
creates the backward nucleon in the $n$th collision via the reaction $R+N\rightarrow N(180^{\degree})+N$. The second mechanism assumes the baryonic resonance formation in $n$ successive collisions with nuclear nucleons, and then its decay into the backward nucleons and pion, $R\rightarrow N(180^{\degree})+\pi $. 
In both considered mechanisms, the largest possible energy of the backward nucleon increases with the number of nuclear nucleons involved in the p~+~$A\rightarrow N(180^{\degree})+\ldots $ reaction. It also noticeably increases with the projectile
proton momentum $p$ up to $p\sim 10$~GeV/$c$ and then goes gradually to its limiting value at $p\rightarrow \infty$. The largest energies of the backward nucleon appear to be close to each other in both discussed mechanisms and coincide  at $p\rightarrow \infty$. However, the masses and longitudinal velocities of baryonic resonances in these two mechanisms are rather different. 

Different aspects of the backward proton production in p~+~$A$ reactions are also studied within the UrQMD simulations. The reactions p~+~$^4$He, p~+~$^{12}$C, and p~+~$^{208}$Pb at the projectile momenta $p=6.9$~GeV/$c$ and $p=158$~GeV/$c$, energy at Dubna Synchrophasotron and top CERN SPS energy, respectively, are considered. 
The energy spectrum of the backward protons  behaves as $\sim A^\alpha$, where $\alpha\cong 2.46$ for light nuclei and $\alpha \cong 0.67$ for heavy nuclei.
We have also found that the maximal energy of the backward proton increases strongly with $A$. 

The UrQMD results  show that the decays $R\rightarrow N(180^{\degree})+\pi$ and the reactions $R+N\rightarrow N(180^{\degree})$ dominate in the UrQMD production of the backward nucleons with large kinetic energy. The second reaction appears to be the main source of the backward nucleons in p~+~Pb reactions. This is qualitatively different from the pion backward production considered in Ref.~\cite{Motornenko:2016sfg} where only resonance decays $R\rightarrow \pi(180^{\degree})+N$ are permitted.

Short range $p$-$n$ correlations inside the nuclei are not implemented  in the present version of the UrQMD model. Their implementation would result in a strong increase of particle  production in the kinematic regions forbidden in p~+~p reactions due to the quasi-elastic collisions of the projectile with nuclear target nucleons.  Considered simultaneously both the short range correlations effects and inelastic re-scatterings of heavy hadron-like states in nuclei are expected to be the competitive mechanisms for a production of final hadrons with momenta forbidden in p~+~p reactions and for the sub-threshold hadron production in low energy p~+~$A$ and $A+A$ collisions. We hope that further experimental studies of p~+~$A$ reactions allow to search for the new heavy hadron-like states and an extension of the hadron mass spectrum to higher mass values. This research can be done by the NA61/SHINE collaboration at the SPS CERN as well as in GSI-HADES, and future FAIR-CBM and NICA-MPD experiments.

\acknowledgments
The authors are  thankful to M.~Gazdzicki,  M.~Strikman, and G.M. Zinovjev for fruitful discussions.
The work of M.I.G. is supported by the Goal-Oriented Program of
Cooperation between CERN and National Academy of Science
of Ukraine ``Nuclear Matter under Extreme Conditions''
(agreement CC/1-2019, No.0118U005343). H.St. acknowledges the support through the Judah M. Eisenberg Laureatus Chair at Goethe University, and the Walter Greiner Gesellschaft, Frankfurt.

\bibliography{references}

\begin{thebibliography}{45}
\expandafter\ifx\csname natexlab\endcsname\relax\def\natexlab#1{#1}\fi
\expandafter\ifx\csname bibnamefont\endcsname\relax
  \def\bibnamefont#1{#1}\fi
\expandafter\ifx\csname bibfnamefont\endcsname\relax
  \def\bibfnamefont#1{#1}\fi
\expandafter\ifx\csname citenamefont\endcsname\relax
  \def\citenamefont#1{#1}\fi
\expandafter\ifx\csname url\endcsname\relax
  \def\url#1{\texttt{#1}}\fi
\expandafter\ifx\csname urlprefix\endcsname\relax\def\urlprefix{URL }\fi
\providecommand{\bibinfo}[2]{#2}
\providecommand{\eprint}[2][]{\url{#2}}

\bibitem[{\citenamefont{Hagedorn}(1965)}]{Hagedorn:1965st}
\bibinfo{author}{\bibfnamefont{R.}~\bibnamefont{Hagedorn}},
  \bibinfo{journal}{Nuovo Cim. Suppl.} \textbf{\bibinfo{volume}{3}},
  \bibinfo{pages}{147} (\bibinfo{year}{1965}).

\bibitem[{\citenamefont{Tanabashi et~al.}(2018)}]{Tanabashi:2018oca}
\bibinfo{author}{\bibfnamefont{M.}~\bibnamefont{Tanabashi}}
  \bibnamefont{et~al.} (\bibinfo{collaboration}{Particle Data Group}),
  \bibinfo{journal}{Phys. Rev.} \textbf{\bibinfo{volume}{D98}},
  \bibinfo{pages}{030001} (\bibinfo{year}{2018}).

\bibitem[{\citenamefont{Hagedorn and Rafelski}(1980)}]{Hagedorn:1980kb}
\bibinfo{author}{\bibfnamefont{R.}~\bibnamefont{Hagedorn}} \bibnamefont{and}
  \bibinfo{author}{\bibfnamefont{J.}~\bibnamefont{Rafelski}},
  \bibinfo{journal}{Phys. Lett.} \textbf{\bibinfo{volume}{97B}},
  \bibinfo{pages}{136} (\bibinfo{year}{1980}).

\bibitem[{\citenamefont{Gorenstein et~al.}(1981)\citenamefont{Gorenstein,
  Petrov, and Zinovev}}]{Gorenstein:1981fa}
\bibinfo{author}{\bibfnamefont{M.~I.} \bibnamefont{Gorenstein}},
  \bibinfo{author}{\bibfnamefont{V.~K.} \bibnamefont{Petrov}},
  \bibnamefont{and} \bibinfo{author}{\bibfnamefont{G.~M.}
  \bibnamefont{Zinovev}}, \bibinfo{journal}{Phys. Lett.}
  \textbf{\bibinfo{volume}{106B}}, \bibinfo{pages}{327} (\bibinfo{year}{1981}).

\bibitem[{\citenamefont{Stoecker et~al.}(1981)\citenamefont{Stoecker, Ogloblin,
  and Greiner}}]{Stoecker:1981za}
\bibinfo{author}{\bibfnamefont{H.}~\bibnamefont{Stoecker}},
  \bibinfo{author}{\bibfnamefont{A.~A.} \bibnamefont{Ogloblin}},
  \bibnamefont{and} \bibinfo{author}{\bibfnamefont{W.}~\bibnamefont{Greiner}},
  \bibinfo{journal}{Z. Phys.} \textbf{\bibinfo{volume}{A303}},
  \bibinfo{pages}{259} (\bibinfo{year}{1981}).

\bibitem[{\citenamefont{Vovchenko et~al.}(2019)\citenamefont{Vovchenko,
  Gorenstein, Greiner, and Stoecker}}]{Vovchenko:2018eod}
\bibinfo{author}{\bibfnamefont{V.}~\bibnamefont{Vovchenko}},
  \bibinfo{author}{\bibfnamefont{M.~I.} \bibnamefont{Gorenstein}},
  \bibinfo{author}{\bibfnamefont{C.}~\bibnamefont{Greiner}}, \bibnamefont{and}
  \bibinfo{author}{\bibfnamefont{H.}~\bibnamefont{Stoecker}},
  \bibinfo{journal}{Phys. Rev.} \textbf{\bibinfo{volume}{C99}},
  \bibinfo{pages}{045204} (\bibinfo{year}{2019}), \eprint{1811.05737}.

\bibitem[{\citenamefont{Bass et~al.}(1998)}]{Bass:1998ca}
\bibinfo{author}{\bibfnamefont{S.~A.} \bibnamefont{Bass}} \bibnamefont{et~al.},
  \bibinfo{journal}{Prog. Part. Nucl. Phys.} \textbf{\bibinfo{volume}{41}},
  \bibinfo{pages}{255} (\bibinfo{year}{1998}), \bibinfo{note}{[Prog. Part.
  Nucl. Phys.41,225(1998)]}, \eprint{nucl-th/9803035}.

\bibitem[{\citenamefont{Bleicher et~al.}(1999)}]{Bleicher:1999xi}
\bibinfo{author}{\bibfnamefont{M.}~\bibnamefont{Bleicher}}
  \bibnamefont{et~al.}, \bibinfo{journal}{J. Phys.}
  \textbf{\bibinfo{volume}{G25}}, \bibinfo{pages}{1859} (\bibinfo{year}{1999}),
  \eprint{hep-ph/9909407}.

\bibitem[{\citenamefont{Cassing and Bratkovskaya}(2008)}]{Cassing:2008sv}
\bibinfo{author}{\bibfnamefont{W.}~\bibnamefont{Cassing}} \bibnamefont{and}
  \bibinfo{author}{\bibfnamefont{E.~L.} \bibnamefont{Bratkovskaya}},
  \bibinfo{journal}{Phys. Rev.} \textbf{\bibinfo{volume}{C78}},
  \bibinfo{pages}{034919} (\bibinfo{year}{2008}), \eprint{0808.0022}.

\bibitem[{\citenamefont{Cassing and Bratkovskaya}(2009)}]{Cassing:2009vt}
\bibinfo{author}{\bibfnamefont{W.}~\bibnamefont{Cassing}} \bibnamefont{and}
  \bibinfo{author}{\bibfnamefont{E.~L.} \bibnamefont{Bratkovskaya}},
  \bibinfo{journal}{Nucl. Phys.} \textbf{\bibinfo{volume}{A831}},
  \bibinfo{pages}{215} (\bibinfo{year}{2009}), \eprint{0907.5331}.

\bibitem[{\citenamefont{Belkacem et~al.}(1998)}]{Belkacem:1998gy}
\bibinfo{author}{\bibfnamefont{M.}~\bibnamefont{Belkacem}}
  \bibnamefont{et~al.}, \bibinfo{journal}{Phys. Rev.}
  \textbf{\bibinfo{volume}{C58}}, \bibinfo{pages}{1727} (\bibinfo{year}{1998}),
  \eprint{nucl-th/9804058}.

\bibitem[{\citenamefont{Andersson et~al.}(1983)\citenamefont{Andersson,
  Gustafson, Ingelman, and Sjostrand}}]{Andersson:1983ia}
\bibinfo{author}{\bibfnamefont{B.}~\bibnamefont{Andersson}},
  \bibinfo{author}{\bibfnamefont{G.}~\bibnamefont{Gustafson}},
  \bibinfo{author}{\bibfnamefont{G.}~\bibnamefont{Ingelman}}, \bibnamefont{and}
  \bibinfo{author}{\bibfnamefont{T.}~\bibnamefont{Sjostrand}},
  \bibinfo{journal}{Phys. Rept.} \textbf{\bibinfo{volume}{97}},
  \bibinfo{pages}{31} (\bibinfo{year}{1983}).

\bibitem[{\citenamefont{Beitel et~al.}(2014)\citenamefont{Beitel, Gallmeister,
  and Greiner}}]{Beitel:2014kza}
\bibinfo{author}{\bibfnamefont{M.}~\bibnamefont{Beitel}},
  \bibinfo{author}{\bibfnamefont{K.}~\bibnamefont{Gallmeister}},
  \bibnamefont{and} \bibinfo{author}{\bibfnamefont{C.}~\bibnamefont{Greiner}},
  \bibinfo{journal}{Phys. Rev.} \textbf{\bibinfo{volume}{C90}},
  \bibinfo{pages}{045203} (\bibinfo{year}{2014}), \eprint{1402.1458}.

\bibitem[{\citenamefont{Baldin et~al.}(1973)}]{Baldin:1973pt}
\bibinfo{author}{\bibfnamefont{A.~M.} \bibnamefont{Baldin}}
  \bibnamefont{et~al.}, \bibinfo{journal}{Yad. Fiz.}
  \textbf{\bibinfo{volume}{18}}, \bibinfo{pages}{79} (\bibinfo{year}{1973}).

\bibitem[{\citenamefont{Baldin et~al.}(1975)}]{Baldin:1974sh}
\bibinfo{author}{\bibfnamefont{A.~M.} \bibnamefont{Baldin}}
  \bibnamefont{et~al.}, \bibinfo{journal}{Sov. J. Nucl. Phys.}
  \textbf{\bibinfo{volume}{20}}, \bibinfo{pages}{629} (\bibinfo{year}{1975}),
  \bibinfo{note}{[Yad. Fiz.20,1201(1974)]}.

\bibitem[{\citenamefont{Amado and Woloshyn}(1976)}]{Amado:1976kn}
\bibinfo{author}{\bibfnamefont{R.~D.} \bibnamefont{Amado}} \bibnamefont{and}
  \bibinfo{author}{\bibfnamefont{R.~M.} \bibnamefont{Woloshyn}},
  \bibinfo{journal}{Phys. Rev. Lett.} \textbf{\bibinfo{volume}{36}},
  \bibinfo{pages}{1435} (\bibinfo{year}{1976}).

\bibitem[{\citenamefont{Frankel}(1977)}]{Frankel:1976bq}
\bibinfo{author}{\bibfnamefont{S.}~\bibnamefont{Frankel}},
  \bibinfo{journal}{Phys. Rev. Lett.} \textbf{\bibinfo{volume}{38}},
  \bibinfo{pages}{1338} (\bibinfo{year}{1977}).

\bibitem[{\citenamefont{Frankfurt and Strikman}(1977)}]{Frankfurt:1977np}
\bibinfo{author}{\bibfnamefont{L.~L.} \bibnamefont{Frankfurt}}
  \bibnamefont{and} \bibinfo{author}{\bibfnamefont{M.~I.}
  \bibnamefont{Strikman}}, \bibinfo{journal}{Phys. Lett.}
  \textbf{\bibinfo{volume}{69B}}, \bibinfo{pages}{93} (\bibinfo{year}{1977}).

\bibitem[{\citenamefont{Frankfurt and Strikman}(1981)}]{Frankfurt:1981mk}
\bibinfo{author}{\bibfnamefont{L.~L.} \bibnamefont{Frankfurt}}
  \bibnamefont{and} \bibinfo{author}{\bibfnamefont{M.~I.}
  \bibnamefont{Strikman}}, \bibinfo{journal}{Phys. Rept.}
  \textbf{\bibinfo{volume}{76}}, \bibinfo{pages}{215} (\bibinfo{year}{1981}).

\bibitem[{\citenamefont{Frankfurt and Strikman}(1988)}]{Frankfurt:1988nt}
\bibinfo{author}{\bibfnamefont{L.~L.} \bibnamefont{Frankfurt}}
  \bibnamefont{and} \bibinfo{author}{\bibfnamefont{M.~I.}
  \bibnamefont{Strikman}}, \bibinfo{journal}{Phys. Rept.}
  \textbf{\bibinfo{volume}{160}}, \bibinfo{pages}{235} (\bibinfo{year}{1988}).

\bibitem[{\citenamefont{Frankfurt et~al.}(2008)\citenamefont{Frankfurt,
  Sargsian, and Strikman}}]{Frankfurt:2008zv}
\bibinfo{author}{\bibfnamefont{L.}~\bibnamefont{Frankfurt}},
  \bibinfo{author}{\bibfnamefont{M.}~\bibnamefont{Sargsian}}, \bibnamefont{and}
  \bibinfo{author}{\bibfnamefont{M.}~\bibnamefont{Strikman}},
  \bibinfo{journal}{Int. J. Mod. Phys.} \textbf{\bibinfo{volume}{A23}},
  \bibinfo{pages}{2991} (\bibinfo{year}{2008}), \eprint{0806.4412}.

\bibitem[{\citenamefont{Baldin}(1977)}]{Baldin:1978bw}
\bibinfo{author}{\bibfnamefont{A.~M.} \bibnamefont{Baldin}},
  \bibinfo{journal}{Fiz. Elem. Chast. Atom. Yadra}
  \textbf{\bibinfo{volume}{8}}, \bibinfo{pages}{429} (\bibinfo{year}{1977}),
  \bibinfo{note}{[,296(1978)]}.

\bibitem[{\citenamefont{Burov et~al.}(1977)\citenamefont{Burov, Lukyanov, and
  Titov}}]{Burov:1976xd}
\bibinfo{author}{\bibfnamefont{V.~V.} \bibnamefont{Burov}},
  \bibinfo{author}{\bibfnamefont{V.~K.} \bibnamefont{Lukyanov}},
  \bibnamefont{and} \bibinfo{author}{\bibfnamefont{A.~I.} \bibnamefont{Titov}},
  \bibinfo{journal}{Phys. Lett.} \textbf{\bibinfo{volume}{67B}},
  \bibinfo{pages}{46} (\bibinfo{year}{1977}).

\bibitem[{\citenamefont{Efremov}(1976)}]{Efremov:1976xw}
\bibinfo{author}{\bibfnamefont{A.~V.} \bibnamefont{Efremov}},
  \bibinfo{journal}{Sov. J. Nucl. Phys.} \textbf{\bibinfo{volume}{24}},
  \bibinfo{pages}{633} (\bibinfo{year}{1976}), \bibinfo{note}{[Yad.
  Fiz.24,1208(1976)]}.

\bibitem[{\citenamefont{Efremov et~al.}(1988)\citenamefont{Efremov, Kaidalov,
  Kim, Lykasov, and Slavin}}]{Efremov:1987mx}
\bibinfo{author}{\bibfnamefont{A.~V.} \bibnamefont{Efremov}},
  \bibinfo{author}{\bibfnamefont{A.~B.} \bibnamefont{Kaidalov}},
  \bibinfo{author}{\bibfnamefont{V.~T.} \bibnamefont{Kim}},
  \bibinfo{author}{\bibfnamefont{G.~I.} \bibnamefont{Lykasov}},
  \bibnamefont{and} \bibinfo{author}{\bibfnamefont{N.~V.}
  \bibnamefont{Slavin}}, \bibinfo{journal}{Sov. J. Nucl. Phys.}
  \textbf{\bibinfo{volume}{47}}, \bibinfo{pages}{868} (\bibinfo{year}{1988}),
  \bibinfo{note}{[Yad. Fiz.47,1364(1988)]}.

\bibitem[{\citenamefont{Braun and Vechernin}(2001)}]{Braun:2001ru}
\bibinfo{author}{\bibfnamefont{M.~A.} \bibnamefont{Braun}} \bibnamefont{and}
  \bibinfo{author}{\bibfnamefont{V.~V.} \bibnamefont{Vechernin}},
  \bibinfo{journal}{Nucl. Phys. Proc. Suppl.} \textbf{\bibinfo{volume}{92}},
  \bibinfo{pages}{156} (\bibinfo{year}{2001}), \bibinfo{note}{[,156(2001)]}.

\bibitem[{\citenamefont{Gorenstein and Zinovev}(1977)}]{Gorenstein:1976zg}
\bibinfo{author}{\bibfnamefont{M.~I.} \bibnamefont{Gorenstein}}
  \bibnamefont{and} \bibinfo{author}{\bibfnamefont{G.~M.}
  \bibnamefont{Zinovev}}, \bibinfo{journal}{Phys. Lett.}
  \textbf{\bibinfo{volume}{67B}}, \bibinfo{pages}{100} (\bibinfo{year}{1977}).

\bibitem[{\citenamefont{Gorenstein et~al.}(1977)\citenamefont{Gorenstein,
  Zinovev, and Shelest}}]{Gorenstein:1977hm}
\bibinfo{author}{\bibfnamefont{M.~I.} \bibnamefont{Gorenstein}},
  \bibinfo{author}{\bibfnamefont{G.~M.} \bibnamefont{Zinovev}},
  \bibnamefont{and} \bibinfo{author}{\bibfnamefont{V.~P.}
  \bibnamefont{Shelest}}, \bibinfo{journal}{Yad. Fiz.}
  \textbf{\bibinfo{volume}{26}}, \bibinfo{pages}{788} (\bibinfo{year}{1977}).

\bibitem[{\citenamefont{Bogatskaya et~al.}(1978)\citenamefont{Bogatskaya,
  Gorenstein, and Zinovev}}]{Bogatskaya:1978jh}
\bibinfo{author}{\bibfnamefont{I.~G.} \bibnamefont{Bogatskaya}},
  \bibinfo{author}{\bibfnamefont{M.~I.} \bibnamefont{Gorenstein}},
  \bibnamefont{and} \bibinfo{author}{\bibfnamefont{G.~M.}
  \bibnamefont{Zinovev}}, \bibinfo{journal}{Yad. Fiz.}
  \textbf{\bibinfo{volume}{27}}, \bibinfo{pages}{856} (\bibinfo{year}{1978}).

\bibitem[{\citenamefont{Bogatskaya et~al.}(1980)\citenamefont{Bogatskaya, Chiu,
  Gorenstein, and Zinovev}}]{Bogatskaya:1979qz}
\bibinfo{author}{\bibfnamefont{I.~G.} \bibnamefont{Bogatskaya}},
  \bibinfo{author}{\bibfnamefont{C.~B.} \bibnamefont{Chiu}},
  \bibinfo{author}{\bibfnamefont{M.~I.} \bibnamefont{Gorenstein}},
  \bibnamefont{and} \bibinfo{author}{\bibfnamefont{G.~M.}
  \bibnamefont{Zinovev}}, \bibinfo{journal}{Phys. Rev.}
  \textbf{\bibinfo{volume}{C22}}, \bibinfo{pages}{209} (\bibinfo{year}{1980}).

\bibitem[{\citenamefont{Anchishkin et~al.}(1982)\citenamefont{Anchishkin,
  Gorenstein, and Zinovev}}]{Anchishkin:1981fb}
\bibinfo{author}{\bibfnamefont{D.~V.} \bibnamefont{Anchishkin}},
  \bibinfo{author}{\bibfnamefont{M.~I.} \bibnamefont{Gorenstein}},
  \bibnamefont{and} \bibinfo{author}{\bibfnamefont{G.~M.}
  \bibnamefont{Zinovev}}, \bibinfo{journal}{Phys. Lett.}
  \textbf{\bibinfo{volume}{108B}}, \bibinfo{pages}{47} (\bibinfo{year}{1982}).

\bibitem[{\citenamefont{Golubyatnikova
  et~al.}(1984)\citenamefont{Golubyatnikova, Shakhanova, and
  Shmonin}}]{Golubyatnikova:1984yh}
\bibinfo{author}{\bibfnamefont{E.~S.} \bibnamefont{Golubyatnikova}},
  \bibinfo{author}{\bibfnamefont{G.~A.} \bibnamefont{Shakhanova}},
  \bibnamefont{and} \bibinfo{author}{\bibfnamefont{V.~L.}
  \bibnamefont{Shmonin}}, \bibinfo{journal}{Acta Phys. Polon.}
  \textbf{\bibinfo{volume}{B15}}, \bibinfo{pages}{585} (\bibinfo{year}{1984}).

\bibitem[{\citenamefont{Kalinkin and Shmonin}(1990)}]{Kalinkin:1989wr}
\bibinfo{author}{\bibfnamefont{B.~N.} \bibnamefont{Kalinkin}} \bibnamefont{and}
  \bibinfo{author}{\bibfnamefont{V.~L.} \bibnamefont{Shmonin}},
  \bibinfo{journal}{Phys. Scripta} \textbf{\bibinfo{volume}{42}},
  \bibinfo{pages}{393} (\bibinfo{year}{1990}).

\bibitem[{\citenamefont{Motornenko and
  Gorenstein}(2017{\natexlab{a}})}]{Motornenko:2016sfg}
\bibinfo{author}{\bibfnamefont{A.}~\bibnamefont{Motornenko}} \bibnamefont{and}
  \bibinfo{author}{\bibfnamefont{M.~I.} \bibnamefont{Gorenstein}},
  \bibinfo{journal}{J. Phys.} \textbf{\bibinfo{volume}{G44}},
  \bibinfo{pages}{025105} (\bibinfo{year}{2017}{\natexlab{a}}),
  \eprint{1604.04308}.

\bibitem[{\citenamefont{Motornenko and
  Gorenstein}(2017{\natexlab{b}})}]{Motornenko:2016uzi}
\bibinfo{author}{\bibfnamefont{A.}~\bibnamefont{Motornenko}} \bibnamefont{and}
  \bibinfo{author}{\bibfnamefont{M.~I.} \bibnamefont{Gorenstein}},
  \bibinfo{journal}{Acta Phys. Polon. Supp.} \textbf{\bibinfo{volume}{10}},
  \bibinfo{pages}{681} (\bibinfo{year}{2017}{\natexlab{b}}),
  \eprint{1610.02950}.

\bibitem[{\citenamefont{Aichelin and Ko}(1985)}]{Aichelin:1986ss}
\bibinfo{author}{\bibfnamefont{J.}~\bibnamefont{Aichelin}} \bibnamefont{and}
  \bibinfo{author}{\bibfnamefont{C.~M.} \bibnamefont{Ko}},
  \bibinfo{journal}{Phys. Rev. Lett.} \textbf{\bibinfo{volume}{55}},
  \bibinfo{pages}{2661} (\bibinfo{year}{1985}).

\bibitem[{\citenamefont{Mosel}(1991)}]{Mosel:1992rb}
\bibinfo{author}{\bibfnamefont{U.}~\bibnamefont{Mosel}}, \bibinfo{journal}{Ann.
  Rev. Nucl. Part. Sci.} \textbf{\bibinfo{volume}{41}}, \bibinfo{pages}{29}
  (\bibinfo{year}{1991}).

\bibitem[{\citenamefont{Hartnack et~al.}(1994)\citenamefont{Hartnack, Sehn,
  Jaenicke, Stoecker, and Aichelin}}]{Hartnack:1993bq}
\bibinfo{author}{\bibfnamefont{G.}~\bibnamefont{Hartnack}},
  \bibinfo{author}{\bibfnamefont{L.}~\bibnamefont{Sehn}},
  \bibinfo{author}{\bibfnamefont{J.}~\bibnamefont{Jaenicke}},
  \bibinfo{author}{\bibfnamefont{H.}~\bibnamefont{Stoecker}}, \bibnamefont{and}
  \bibinfo{author}{\bibfnamefont{J.}~\bibnamefont{Aichelin}},
  \bibinfo{journal}{Nucl. Phys.} \textbf{\bibinfo{volume}{A580}},
  \bibinfo{pages}{643} (\bibinfo{year}{1994}).

\bibitem[{\citenamefont{Steinheimer and Bleicher}(2016)}]{Steinheimer:2015sha}
\bibinfo{author}{\bibfnamefont{J.}~\bibnamefont{Steinheimer}} \bibnamefont{and}
  \bibinfo{author}{\bibfnamefont{M.}~\bibnamefont{Bleicher}},
  \bibinfo{journal}{J. Phys.} \textbf{\bibinfo{volume}{G43}},
  \bibinfo{pages}{015104} (\bibinfo{year}{2016}), \eprint{1503.07305}.

\bibitem[{\citenamefont{Steinheimer et~al.}(2017)\citenamefont{Steinheimer,
  Botvina, and Bleicher}}]{Steinheimer:2016jjk}
\bibinfo{author}{\bibfnamefont{J.}~\bibnamefont{Steinheimer}},
  \bibinfo{author}{\bibfnamefont{A.}~\bibnamefont{Botvina}}, \bibnamefont{and}
  \bibinfo{author}{\bibfnamefont{M.}~\bibnamefont{Bleicher}},
  \bibinfo{journal}{Phys. Rev.} \textbf{\bibinfo{volume}{C95}},
  \bibinfo{pages}{014911} (\bibinfo{year}{2017}), \eprint{1605.03439}.

\bibitem[{\citenamefont{Gallmeister et~al.}(2018)\citenamefont{Gallmeister,
  Beitel, and Greiner}}]{Gallmeister:2017ths}
\bibinfo{author}{\bibfnamefont{K.}~\bibnamefont{Gallmeister}},
  \bibinfo{author}{\bibfnamefont{M.}~\bibnamefont{Beitel}}, \bibnamefont{and}
  \bibinfo{author}{\bibfnamefont{C.}~\bibnamefont{Greiner}},
  \bibinfo{journal}{Phys. Rev.} \textbf{\bibinfo{volume}{C98}},
  \bibinfo{pages}{024915} (\bibinfo{year}{2018}), \eprint{1712.04018}.

\bibitem[{\citenamefont{Gazdzicki et~al.}(1998)\citenamefont{Gazdzicki,
  Gorenstein, and Mrowczynski}}]{Gazdzicki:1997sg}
\bibinfo{author}{\bibfnamefont{M.}~\bibnamefont{Gazdzicki}},
  \bibinfo{author}{\bibfnamefont{M.~I.} \bibnamefont{Gorenstein}},
  \bibnamefont{and}
  \bibinfo{author}{\bibfnamefont{S.}~\bibnamefont{Mrowczynski}},
  \bibinfo{journal}{Eur. Phys. J.} \textbf{\bibinfo{volume}{C5}},
  \bibinfo{pages}{129} (\bibinfo{year}{1998}), \eprint{nucl-th/9701013}.

\bibitem[{\citenamefont{Weyer}(1990)}]{Weyer:1990ye}
\bibinfo{author}{\bibfnamefont{H.~J.} \bibnamefont{Weyer}},
  \bibinfo{journal}{Phys. Rept.} \textbf{\bibinfo{volume}{195}},
  \bibinfo{pages}{295} (\bibinfo{year}{1990}).

\bibitem[{\citenamefont{Bierlich et~al.}(2018)\citenamefont{Bierlich,
  Gustafson, Lönnblad, and Shah}}]{Bierlich:2018xfw}
\bibinfo{author}{\bibfnamefont{C.}~\bibnamefont{Bierlich}},
  \bibinfo{author}{\bibfnamefont{G.}~\bibnamefont{Gustafson}},
  \bibinfo{author}{\bibfnamefont{L.}~\bibnamefont{Lönnblad}},
  \bibnamefont{and} \bibinfo{author}{\bibfnamefont{H.}~\bibnamefont{Shah}},
  \bibinfo{journal}{JHEP} \textbf{\bibinfo{volume}{10}}, \bibinfo{pages}{134}
  (\bibinfo{year}{2018}), \eprint{1806.10820}.

\bibitem[{\citenamefont{Frankel et~al.}(1979)}]{Frankel:1979uq}
\bibinfo{author}{\bibfnamefont{S.}~\bibnamefont{Frankel}} \bibnamefont{et~al.},
  \bibinfo{journal}{Phys. Rev.} \textbf{\bibinfo{volume}{C20}},
  \bibinfo{pages}{2257} (\bibinfo{year}{1979}).

\end{thebibliography}

\end{document}